\documentclass[11pt]{article}

\usepackage{amsmath, amssymb}
\usepackage{comment}
\usepackage{graphicx}
\usepackage{bm}
\usepackage[version=4]{mhchem}
\usepackage{float}
\usepackage{xcolor}
\usepackage{hyperref}
\usepackage{gensymb}
\usepackage[a4paper, margin=2.5cm]{geometry}
\usepackage{authblk}  
\usepackage[numbers,sort&compress]{natbib}
\usepackage{caption}
\usepackage{setspace}

\DeclareCaptionType{extfigure}[Supplementary Fig.][List of Supplementary Figures]

\usepackage[labelfont=bf]{caption}

\hypersetup{
  colorlinks   = true,
  urlcolor     = blue,
  linkcolor    = blue,  
  citecolor    = blue   
}

\title{\bfseries Electron–phonon coupling revealed by charge density fluctuations in cuprate superconductors}

\author[1]{Martina Fedele\thanks{e-mail: \protect\href{mailto:martina.fedele@polimi.it}{martina.fedele@polimi.it}}}
\author[1,2]{Giacomo Merzoni}
\author[1]{Marco Moretti Sala}
\author[1]{Francesco Rosa}
\author[3]{Nicholas B. Brookes}
\author[4]{Floriana Lombardi}
\author[5]{Sergio Caprara}
\author[1,6]{Giacomo Ghiringhelli\thanks{e-mail: \protect\href{mailto:giacomo.ghiringhelli@polimi.it}{giacomo.ghiringhelli@polimi.it}}}
\author[4,7]{Riccardo Arpaia\thanks{e-mail: \protect\href{mailto:riccardo.arpaia@unive.it}{riccardo.arpaia@unive.it}}}

\affil[1]{Dipartimento di Fisica, Politecnico di Milano, piazza Leonardo da Vinci 32, I-20133 Milano, Italy}
\affil[2]{European XFEL, Holzkoppel 4, D-22869 Schenefeld,  Germany}
\affil[3]{ESRF, The European Synchrotron, 71 Avenue des Martyrs, CS 40220, F-38043 Grenoble, France}
\affil[4]{Quantum Device Physics Laboratory, Department of Microtechnology and Nanoscience, Chalmers University of Technology, SE-41296 Göteborg, Sweden}
\affil[5]{Dipartimento di Fisica, Università di Roma ``La Sapienza'', P.le Aldo Moro 5, I-00185 Roma, Italy}
\affil[6]{CNR-SPIN, Dipartimento di Fisica, Politecnico di Milano, I-20133 Milano, Italy}
\affil[7]{Department of Molecular Sciences and Nanosystems, Ca’ Foscari University of Venice, I-30172 Venice, Italy}

\date{}
\begin{document}
\newcommand{\gm}[1]{\textcolor{magenta}{#1}}
\maketitle
\newpage
\begin{abstract} 
Electron–phonon coupling (EPC) is a key interaction governing lattice dynamics, charge transport, and collective electronic phases in quantum materials. In several families of unconventional superconductors, including transition-metal dichalcogenides and kagome metals, growing evidence points to a cooperative role of EPC and dynamic charge-density fluctuations (CDF) in stabilizing superconductivity. Yet, how the strength of EPC evolves across phase diagrams and relates to superconducting properties in strongly correlated systems remains an open question.

Here we investigate the interplay between phonons and the CDF recently identified in cuprate superconductors. Using resonant inelastic x-ray scattering, we track the dispersion and intensity of bond-stretching phonons in YBa$_2$Cu$_3$O$_{7-\delta}$ over wide ranges of doping, temperature, and momentum. We observe that both the phonon softening at the CDF wave vector and the EPC strength, extracted from a pronounced phonon intensity anomaly, are maximized near $p = 0.19$, where superconducting properties are optimal and CDF intensity is strongest.

These results identify dynamic charge-density fluctuations, rather than quasi-static charge density waves, as the dominant source of phonon renormalization in cuprates, and establish a direct correlation between EPC strength and the superconducting dome. More broadly, our measurements emphasize EPC as a doping dependent property of correlated materials, shaped by the electronic environment in which lattice vibrations are embedded. 
\end{abstract}

\newpage

\section{\label{sec:introduction}Introduction}

Electron–phonon coupling (EPC) is a basic interaction in solids. It links the motion of electrons to lattice vibrations and plays an essential role in many physical phenomena, such as electrical transport, structural phase transitions, and superconductivity. In conventional superconductors, EPC is the key ingredient of the Bardeen–Cooper–Schrieffer (BCS) theory: phonons mediate an effective attraction between electrons, leading to Cooper pairing and dissipationless transport.

The situation becomes more complex in unconventional superconductors. In many quantum materials with strong electron–electron interactions, superconductivity appears together with other forms of electronic order, such as spin or charge density modulations~\cite{castellani1996non, neto2001charge, fradkin2015colloquium, wang2022paramagnons, neupert2022charge}. These additional orders indicate that superconductivity in these systems cannot be understood by EPC alone. Instead, it  seems to emerge from the interplay of multiple degrees of freedom, including electronic correlations and collective fluctuations.

High-temperature cuprate superconductors are a prime example of this complexity. Despite extensive experimental and theoretical efforts, the microscopic mechanism behind their superconductivity remains unresolved~\cite{keimer2015quantum}. Strong electronic correlations are widely considered essential, yet cuprates also show clear signatures of significant electron–phonon coupling~\cite{lanzara2001evidence, kim2002anomalous, shen2004missing, ronning2005anomalous}. This makes it crucial to understand what role EPC plays in these materials, and whether it actively contributes to superconductivity or merely accompanies it.

Recent results from other families of unconventional superconductors suggest a new perspective. In systems such as transition-metal dichalcogenides and kagome metals, superconductivity is promoted by the cooperation between EPC and dynamical charge-density fluctuations (CDF)~\cite{kogar2017observation, hinlopen2024lifshitz, PhysRevB.CDFKagome, kongruengkit2025kagomeCDF}. In this view, EPC is not a fixed interaction, but an emergent property that depends on the electronic environment, and therefore on temperature and doping.

Inspired by this perspective, we investigate whether a similar cooperative mechanism operates in cuprate superconductors. We design an experiment that enables us to follow the evolution of electron–phonon coupling and its connection to superconductivity across the phase diagram as a function of hole doping in YBa$_2$Cu$_3$O$_{7-\delta}$ (YBCO). To this end, we employ Resonant inelastic x-ray scattering (RIXS), a technique that simultaneously probes lattice vibrations and charge excitations~\cite{chaix2017dispersive, li2020multiorbital, lin2020strongly, Huang2021, wang2021charge, lee2021spectroscopic,WSL_PRB_softenining, tam2022charge, scott2023low}. Using Cu $L_3$-edge RIXS, we map the bond-stretching (BS) phonons and the associated charge modulations as a function of momentum, temperature, and doping. We observe a pronounced softening of the BS phonons near a characteristic wave vector $\mathbf{q}_\mathrm{c}$, in agreement with earlier inelastic neutron scattering (INS) results~\cite{PintschoviusTRE, reichardt1994anharmonicity, Chaplot1995, Pintschovius2002,Pintschovius2003,Pintschovius2004,Pintschovius2005,reznik2006electron, Reznik2010}, and we detect charge-density modulations at the same momentum. While this coincidence has often been attributed to quasi-static charge-density waves (CDW)~\cite{GhiringhelliCDW, chang2012direct, comin2014charge, wu2015incipient, hayden2024charge}, our measurements reveal inconsistencies with this assignment. The temperature- and momentum-dependence of the BS-phonon softening does not track that of the CDW peak. Instead, they closely mirror the behavior of the recently identified short-ranged, dynamic CDF~\cite{ArpaiaUNO, yu2020unusual, wang2020doping, wang2020high, arpaia2021charge, von2023fate}, which persist up to room temperature and are most intense near optimal doping ($p\!\approx\!0.19$)~\cite{NatCommRArpaia}. These dynamic fluctuations therefore provide a natural explanation for the observed phonon anomalies.

Our results lead to two central conclusions. First, dynamic CDF control the momentum, temperature, and doping dependence of the BS-phonon renormalization. Second, the effective strength of the electron–phonon coupling—quantified through the phonon softening and peak intensity~\cite{BraicovichPhon, wang2021charge}—closely follows the evolution of the CDF and reaches a maximum at $p\!\approx\!0.19$, where superconductivity is strongest~\cite{bernhard2001anomalous, tallon2003superfluid, tallon2014thermodynamics, grissonnanche2014direct}.

Taken together, these findings establish how electron–phonon coupling evolves within a strongly correlated electronic environment and highlight its active role, through cooperation with dynamic charge-density fluctuations, in high-temperature superconductivity. More broadly, they support a picture in which EPC is not a static interaction, but an emergent property shaped by collective electronic dynamics, pointing toward a unifying mechanism across different families of unconventional superconductors.


This perspective is also consistent with proposals that band-structure and quantum-geometry can enhance EPC~\cite{yu2024non, yu2025quantum}; in particular, kagome metals host flat-bands that naturally amplify EPC-mediated responses~\cite{julku2021quantum}, providing a plausible setting for the observed cooperation between EPC and CDF.

\section{\label{sec: exp res}Experimental results}

We have investigated 100 nm-thick YBCO films (for sample details, see Ref.~\cite{ArpaiaGrowth} and Methods \ref{samples}) with doping levels of $p=0$, $0.06$, $0.13$, $0.19$, and $0.23$, thus spanning most of the phase diagram - from the Mott insulating regime to the overdoped region. YBCO was chosen due to its stronger CDF signal compared to other cuprates, which allows for a clearer separation from CDW, and because the CDF doping dependence has been previously characterized \cite{NatCommRArpaia}. 
RIXS measurements at the Cu $L_3$ edge ($\approx 931$ eV) were performed at the ID32 beamline of ESRF - The European Synchrotron, using an energy resolution of approximately 40 meV. The incident x-rays were $\sigma$-polarized (perpendicular to the scattering plane) to enhance the charge response cross-section~\cite{Moretti,Minola}. The scattering angle was fixed at $2\theta = 149.5^\circ$ throughout the entire experiment. RIXS spectra were acquired across a wide momentum range (0.14–0.48 r.l.u.) along the $(H,0)$ direction and at multiple temperatures from 20~K to room temperature. Additionally, for the samples with $p=0.13$ and $p=0.19$, we performed equivalent momentum scans as a function of the azimuthal angle $\phi$, from the $(H,0)$ direction up to $(H,H)$, thereby covering most of the Brillouin zone between the $\Gamma$, X, and M points. A schematic representation of the RIXS scattering geometry is provided in Supplementary Fig. \ref{fig:RIXSgeometry}. 

In Figure~\ref{fig:figure1}(a)–(b) we show RIXS maps at 20~K for the samples $p=0.13$ and $p=0.19$. An intense and quasi-static peak associated with CDW is observed at $\mathbf{q}_\mathrm{c}\sim0.32$ r.l.u. for $p=0.13$, while a much weaker peak at finite energy is present for $p=0.19$ at a similar momentum and is attributed to CDF. 
To extract the energy position of the BS phonons, a multi-peak fitting procedure was applied to the spectra (see Methods \ref{analysis}), as illustrated in Figure~\ref{fig:figure1}(c)–(f). At both dopings, it is evident that the energy of the BS phonons is lower at momenta close to $\mathbf{q}_\mathrm{c}$ compared to $\mathbf{q}=0.44$ r.l.u, as emphasized by the blue bars. To disentangle the CDW and CDF contributions to the phonon softening, we performed analyses as a function of temperature (see Fig.~\ref{fig:figure2}) and azimuthal angle (see Fig.~\ref{fig:figure3}).

Figure~\ref{fig:figure2}(a)–(b) shows the temperature dependence of the BS phonon energy dispersion for both doping levels, while the corresponding quasi-elastic integrated intensity within the energy range [–100, 35] meV is displayed in Fig.~\ref{fig:figure2}(c)–(d). A pronounced softening of the bond-stretching mode is observed, peaking around $\mathbf{q}_\mathrm{c}$. Notably, we find that the softening (i) is comparable between the two doping levels but slightly more pronounced at $p=0.19$, where CDW are negligible and only CDF are present; (ii) is maximized at 20 K, consistently with CDF behavior but in contrast to CDW, which are suppressed below $T_\mathrm{c}$ \cite{ghiringhelli2012long}; and (iii) monotonically decreases with increasing temperature, yet persists up to the highest investigated temperature in both cases - well above the CDW onset temperature for $p=0.13$.

To quantify the softening, $\Delta E_{p,T}$, we use as a reference the $p \sim 0$ sample, in which no softening of the BS phonon dispersion is observed and the phonon energy remains completely temperature-independent (see Methods \ref{p0}). $\Delta E_{p,T}$ is defined as the difference $\bar{E}_0 - \bar{E}_{p,T}$, where $\bar{E}_{p,T}$ is the arithmetic mean of the phonon energy within the central $\mathbf{q}$-range [0.28-0.35] r.l.u., calculated at a given temperature $T$ and doping level $p$. In this definition, $\bar{E}_0$ denotes the average energy extracted from the dispersion at $p \sim 0$.
Figure~\ref{fig:figure4}(a) displays the integrated intensity of the quasi-elastic peak, obtained by Lorentzian fitting of the $\mathbf{q}$-dependent profiles presented in Fig.~\ref{fig:figure2}(c)-(d). For $p = 0.19$, this intensity is dominated by CDF, which exhibit a mild and monotonic temperature dependence, closely resembling that of $\Delta E_{p,T}$, shown in Figure~\ref{fig:figure4}(b). In contrast, the underdoped sample displays an intense CDW peak that reaches its maximum intensity near $T_\mathbf{c}$ and then rapidly decreases, disappearing above 200 K. The softening decreases linearly with temperature at both dopings, and is stronger at $p = 0.19$. \\
The connection between phonon softening and CDF is further supported by the analysis of the azimuthal angle dependence. In Figure~\ref{fig:figure3}(a)–(b), the phonon softening remains clearly visible up to $\phi = 45^\circ$ at both doping levels, although reduced in magnitude. This behavior reflects the almost isotropic momentum distribution of CDF, whose intensity is maximized along the $(H,0)$ direction and evolves gradually along the azimuthal direction. In contrast, the sharper and more anisotropic CDW signal in the $p=0.13$ sample becomes almost negligible already at $\phi = 15^\circ$ (see Methods \ref{anglescan}).

Finally, in Figure~\ref{fig:figure4}(c), we present the BS phonon energy at $T=20$~K as a function of both doping $p$ and momentum $\mathbf{q}$.  A pronounced dip appears at $\mathbf{q}_\mathrm{c}$ and $p=0.19$, which becomes less marked at both lower and higher doping levels, yet remains finite even at extreme values such as $p=0.06$ and $p=0.23$, where CDW are absent.  When plotting $\Delta E$ as a function of doping, we find that it closely follows the doping dependence previously reported for the CDF intensity \cite{NatCommRArpaia} (see Fig.~\ref{fig:figure4}(d)), further enriching the complex landscape emerging around the critical doping $p\sim0.19$. 

\textit{Taken together, our combined and systematic investigation of bond-stretching phonons and charge order (i.e., CDW and CDF) versus temperature, momentum, and doping, reveals a strong connection between phonon softening and charge density fluctuations.} This interpretation is further supported by recent reports showing that uniaxial strain strongly modulates the CDW peak—making it highly anisotropic—while leaving the BS phonon softening unaffected~\cite{Martinelli2024}. Such strain independence aligns with the known insensitivity of CDF to strain, which arises from their short-range nature~\cite{wahlberg2021restored, boyle2021large}. We emphasize that our observation of the doping dependence of BS phonon softening is consistent with previous observations from inelastic neutron scattering experiments in YBCO ~\cite{Chaplot1995, PintschoviusTRE,Pintschovius2002,Pintschovius2003,Pintschovius2004,Pintschovius2005,reznik2006electron,Reznik2010}, which reported enhanced phonon softening upon lowering the temperature and increasing the oxygen doping, reaching a maximum near $p\sim0.19$ (see grey symbols in Figure~\ref{fig:figure2}(a)–(b)) \footnote{The main difference between previous INS measurements and our current RIXS data lies in the twinning state of the YBCO samples. The YBCO crystals studied by INS were untwinned, resulting in different amount of BS phonon softening along the $a$ and $b$ crystallographic directions. In contrast, our thin films are twinned, and the effect of CDF on the phonons becomes effectively isotropic between $a$ and $b$, with the same softening observed along both directions. In Figure~\ref{fig:figure2}, we have compared our data with the dispersion measured by INS along the $b$-axis, where a more pronounced softening  was observed.
This is an approximation that nonetheless yields reasonably good quantitative agreement. This observation highlights the need for a dedicated investigation into the possible anisotropy of CDF between the $a$ and $b$ directions in untwinned films (e.g., grown on vicinal-angle STO substrates), which will be addressed in future work.}.

Beyond the evidence for coupling between CDF and phonons, and their entanglement with superconductivity, our data also shed new light on the role of the EPC itself. Our RIXS measurements allow us to track the temperature, doping, and momentum dependence of the intensity associated with the BS phonons, which provides direct insight into the strength of the underlying EPC. Figure~\ref{fig:figure4}(e) displays the momentum dependence of the BS phonon intensity across different doping levels. A clear doping dependence emerges: starting from the undoped, insulating sample, where the intensity follows the expected $\sin^2(\pi q)$ behavior, the signal increases by more than a factor of two, reaching a maximum at $p = 0.19$, before decreasing again at higher doping. When averaging the BS intensity over the high-$\mathbf{q}$ region ($q > 0.3$), where uncertainties are minimized, we obtain a quantity $I_{\mathrm{BS}}$ that mirrors the doping dependence of both the CDF intensity and the phonon softening (Figure~\ref{fig:figure4}(f)). This correlation is also confirmed by the temperature dependence of the breathing phonon intensity at $p=0.13$ and $p=0.19$ (see Methods~\ref{BSInt} and Supplementary Figure~\ref{fig:IntBS}), which again closely follows that of the CDF signal and the softening amplitude. Taken together, these results show that the EPC in cuprates: (i) is strongly doping dependent, (ii) it is intimately connected to the presence of CDF, sharing the same doping trend, and (iii) it is maximized at $p \sim 0.19$, where superconductivity is most robust - pointing to an active role of the EPC, possibly entangled with CDF, in shaping the superconducting state.

\section{\label{sec: discussion}Discussion}

In conventional Peierls-like pictures of CDW formation, EPC drives a complete softening of a phonon branch at the ordering wavevector, ultimately leading to a static lattice distortion. 

Although partial, the softening of the bond-stretching mode reported in cuprates by inelastic neutron scattering was naturally interpreted as an indirect fingerprint of a strong EPC to an underlying charge instability ~\cite{PintschoviusTRE, reichardt1994anharmonicity, Chaplot1995, Pintschovius2002,Pintschovius2003,Pintschovius2004,Pintschovius2005,reznik2006electron, Reznik2010}. In parallel, ARPES provided direct and independent evidence for such a strong EPC in cuprates ~\cite{lanzara2001evidence, kim2002anomalous, shen2004missing, ronning2005anomalous}, but lacks sensitivity to the specific phonon modes. Here we have exploited the unique capabilities of RIXS to simultaneously probe charge and lattice degrees of freedom and provided direct evidence of strong coupling between the bond-stretching phonon mode and CDF.

A direct coupling between charge modulations and phonon anomalies has been theoretically discussed in terms of plasmon--phonon interaction mechanisms~\cite{Castellani,becca1996charge} and Fano interference between charge and lattice dynamics~\cite{chaix2017dispersive}. Can we thus associate the phonon renormalization observed here to CDF rather than to CDW? On the one hand, multiple experimental probes—including inelastic X-ray scattering, nuclear magnetic resonance, and X-ray photon correlation spectroscopy—indicate that CDW are essentially static in nature~\cite{wu2015incipient, blackburn2013inelastic, chen2016remarkable,thampy2017static}. Such a quasi-static character is difficult to reconcile with the partial phonon softening observed experimentally. On the other hand, even if CDW retain a residual dynamics (with characteristic energies below the meV scale), their total spatial extent remains too limited to induce an appreciable renormalization. Indeed, we have previously shown~\cite{ArpaiaUNO} that the total volume of the broad-in-$\mathbf{q}$ peak associated with CDF is substantially larger than that of the much narrower CDW peak (see Methods~\ref{volumes}). 

The temperature dependence further supports this conclusion. To interpret the evolution of the phonon softening $\Delta E$ in terms of CDF, it is essential to consider the CDF characteristic energy $\omega_0$. This quantity is directly obtained from the RIXS spectra as the energy position of the Gaussian peak associated with CDF (e.g., Fig.~\ref{fig:figure1}(c--f)), and its temperature evolution is reported in Fig.~\ref{fig:figure4}(b). Whether determined directly or indirectly---by extrapolating it from the FWHM of the CDF peak in momentum space (e.g., Fig.~\ref{fig:figure2}(d))---$\omega_0$ increases monotonically with temperature, closely tracking the concurrent decrease of $\Delta E$ (Methods~\ref{energyCDF}). This correspondence demonstrates that the relevant control parameter for the phonon renormalization is not the mere presence of charge correlations, but the reduction of their characteristic energy scale $\omega_0$ toward zero.

A similarly sharp message emerges from the doping dependence. Both the magnitude of the phonon softening and the strength of the CDF follow a dome-like dependence on doping, closely mirroring the superconducting dome of the cuprates. They peak at $p\sim 0.19$, just beyond optimal doping, where several key indicators of superconducting robustness---including the superfluid density, critical current density, and upper critical field---reach their maximum values~\cite{bernhard2001anomalous, tallon2003superfluid, tallon2014thermodynamics, grissonnanche2014direct}. Importantly, in our data the strongest phonon renormalization coincides with the doping where the CDF energy scale is minimal. Moreover, the RIXS spectral weight of the bond-stretching phonons---expected to scale with the effective EPC---also maximizes around $p\sim 0.19$. This closes the loop: the bond-stretching softening, together with the EPC-sensitive phonon response, is enhanced precisely when the dynamic charge correlations are closest to a zero-energy instability. This alignment suggests that electron--phonon interaction may play a central role in setting the superconducting energy scale~\cite{perali1996d,lanzara2001evidence, gweon2004unusual, jiang2024correlation}, and demonstrates that the degree of phonon renormalization is tuned primarily by the low-energy charge dynamics. 

As the CDF energy decreases, either due to temperature or doping, the EPC is sufficient to drive the system toward a pre-instability, yet without producing the full soft-mode collapse associated with static order. In this sense, dynamic CDF ``steer'' the lattice toward instability while avoiding the frozen distortion expected for quasi-static CDW. This distinction is also conceptually favorable for superconductivity: when quasi-static CDW become prominent, superconductivity is known to be suppressed, consistent with a competition scenario in which the same EPC-enhanced instability can instead support stripe-like order that is incompatible with a robust condensate.

The implications of our findings extend beyond cuprates. Related phenomenology has been discussed in other unconventional superconductors, including transition metal dichalcogenides and kagome metals, where charge fluctuations observed well above the onset of quasi-static CDW~\cite{holt2001x,monney2016revealing, chen2022charge, cao2023competing, liu2024driving, guo2025plane, fragkos2025electron, liu2025fluctuated} have been implicated---together with EPC---in mediating $s$-wave pairing~\cite{kogar2017observation, hinlopen2024lifshitz,PhysRevB.CDFKagome, kongruengkit2025kagomeCDF}, potentially aided by band-structure and quantum-geometry effects that enhance EPC, e.g., in the presence of flat-bands in kagome systems \cite{julku2021quantum}. In cuprates, the emergence of $d$-wave superconductivity from a Mott-insulating parent state adds an additional layer of complexity, and nearly antiferromagnetic spin fluctuations may contribute, for example by favoring the $d$-wave pairing channel. Nevertheless, our results show that the phonon renormalization and the strength of the dynamic charge-density fluctuations evolve in lockstep across doping, and both follow the superconducting dome.
 This strengthens a scenario in which charge dynamics, lattice interactions, and superconductivity are deeply intertwined---and motivates renewed attention to the joint role of CDF and EPC as active ingredients of pairing in cuprates.

\section{\label{sec:methods}Methods}
\subsection{Samples}\label{samples}
Four YBa$_2$Cu$_3$O$_{7-\delta}$ (YBCO) thin films and one Y$_{0.7}$Ca$_{0.3}$Ba$_2$Cu$_3$O$_{7-\delta}$ (Ca-YBCO) thin film, each with a thickness of $t = 100$~nm, were deposited by pulsed laser deposition on 5~$\times$~5~mm$^2$ (001)-oriented SrTiO$_3$ substrates. Details of the growth procedure are provided in \cite{ArpaiaGrowth}.

The corresponding zero-resistance critical temperatures ($T_c$) are 0~K, 12~K, 65~K, and 86~K for the YBCO films, and 56~K for the Ca-YBCO film. These values were achieved through post-annealing under varying oxygen partial pressures: 4.9~$\times$~10$^{-5}$, 1.1~$\times$~10$^{-4}$, 1.2~$\times$~10$^{-2}$, and 6.5~$\times$~10$^{2}$~Torr for the YBCO films, and 6.5~$\times$~10$^{2}$~Torr for the Ca-YBCO film.

The hole doping levels $p = 0$, $0.06$, $0.13$, and $0.19$ for the YBCO films and $p = 0.23$ for the Ca-YBCO film were estimated using a method previously validated for single crystals \cite{liang2006evaluation}, based on the measured $T_c$ and the $c$-axis lattice parameter determined by X-ray diffraction.
\subsection{Data analysis}\label{analysis}
To enable a direct comparison between spectra acquired under varying experimental conditions, all RIXS spectra were corrected for self-absorption and normalized to the integral of the inter-orbital $dd$ excitations in the energy loss range [1~eV, 3~eV]. 
Self-absorption coefficients were calculated individually for each momentum transfer, taking into account the experimental geometry, incident energy, and polarization. A detailed description of the self-absorption correction procedure can be found in \cite{NatCommRArpaia}.

Once the RIXS spectra are normalized, the peaks associated with charge order (charge density waves and charge density fluctuations) and the bond-stretching (BS) phonon mode are extracted for each momentum transfer $\mathbf{q}$. As shown in Supplementary Fig. \ref{fig:fitting}, the quasielastic region is modeled as the sum of multiple components. The purely elastic peak (red line in the figure), attributed to the elastic scattering signal originated from surface defects and, in the case of underdoped samples, to CDW, is resolution-limited. In contrast, the full width at half maximum (FWHM) of the low-energy peak (blue line in the figure) associated to CDF is treated as a free fitting parameter. Indeed, it consistently exceeds the energy resolution, in agreement with previous reports~\cite{yu2020unusual}.
Initially, the energy position of the CDF peak was treated as a free fitting parameter. However, it was later fixed to the value near $q_{\mathrm{CDF}}$, where the signal is more reliably determined. This choice is justified by the fact that the CDF intensity is maximized at $q_{\mathrm{CDF}}$,  while at both lower and higher $q$ values the signal is intrinsically weaker and its extraction is further complicated by overlap with the specular reflection and the BS phonon, respectively.
The BS phonon branch is modeled using a Gaussian  (violet shaded peak in the figure), with both its energy position and FWHM treated as free parameters. A second Gaussian is included to account for the phonon overtone (yellow line in the figure), with a free FWHM and an energy  fixed at twice the BS phonon energy. Finally, an antisymmetrized Lorentzian function (green line in the figure) is used to model the spectral contribution of paramagnon excitations.
In the region between the elastic peak and the BS phonon, charge-order-related excitations dominate so no additional peak associated with bond-buckling phonons was required.

\subsection{\textbf{Bond-stretching phonon dispersion in insulating YBCO ($p$ $\sim$ 0)}}\label{p0}
For the undoped sample, RIXS measurements were performed at both low ($T = 20$~K) and high temperature ($T = 200$~K). 
As shown in Supplementary Fig. \ref{fig:Dopdep}, the bond-stretching phonon dispersion remains completely unchanged with momentum and temperature, confirming the absence of any softening effect. This behavior sharply contrasts with the doping-dependent phonon softening observed in all superconducting samples, and indicates that in the undoped Mott insulating regime the phonon energy is independent of both momentum and temperature (within the investigated ranges).

\subsection{\textbf{Azimuthal angle dependence}}\label{anglescan}
The scattering angle was fixed at $2\theta = 149.5^\circ$ throughout the entire experiment. This experimental geometry was optimized to access a transferred momentum of $|\mathbf{q}| = 0.91$~\AA$^{-1}$, which allows full coverage of the first Brillouin zone along the $(H, 0)$ direction ($0.5$~r.l.u.~$\approx 0.81$~\AA$^{-1}$). 

For the azimuthal angle dependence, $\phi$ was varied up to 45° to scan reciprocal space from the $(H, 0)$ direction up to the $(H, H)$ direction, thus spanning a large portion of the Brillouin zone between the $\Gamma$, $X$, and $M$ points.

A schematic representation of the RIXS scattering geometry, including the definition of the azimuthal angle $\phi$, is shown in Supplementary Fig.~\ref{fig:RIXSgeometry}.

To complement the maps presented in Figure~\ref{fig:figure3} of the main text, the underlying bond-stretching (BS) phonon dispersions are shown in Supplementary Fig.~\ref{fig:Phiscan}(a)-(b) as a function of the azimuthal angle $\phi$ for both doping levels. These data provide a clearer view of the phonon energy evolution across the wedge-shaped region of the Brillouin zone connecting the $\Gamma$–X and $\Gamma$–M directions. For direct comparison, the corresponding integrated quasielastic intensities in the energy range [$-35$, 100]~meV are shown in panels (c) and (d).

\subsection{\textbf{Bond-stretching phonon intensity}}\label{BSInt}

For a comprehensive characterization of the phonon behavior, we also report the intensity of the bond-stretching phonon peak as a function of temperature (Supplementary Fig.~\ref{fig:IntBS}(a)--(b)) and azimuthal angle (Supplementary Fig.~\ref{fig:IntBS}(c)--(d)). In an ideal, non-interacting scenario, the momentum dependence of the phonon intensity is expected to follow a $\sin^2(\pi q)$ behavior~\cite{BraicovichPhon}. 
In our case, vice versa, a pronounced anomaly emerges at low temperature and near $\phi = 0^{\circ}$: for momenta beyond $\mathbf{q}_{\mathrm{c}}$ the phonon intensity rises markedly above the $\sin^{2}(\pi q)$ trend, with this excess gradually diminishing - though not vanishing - as either temperature or azimuthal angle increases.


This anomaly is consistent with a coupling/interference between the bond-stretching phonon and charge order, as proposed in Refs. \cite{chaix2017dispersive, lin2020strongly, wang2021charge, lee2021spectroscopic}. In our data, its dependence on doping, temperature, and azimuthal angle closely tracks that of charge-density fluctuations: the excess phonon intensity above the $\sin^{2}(\pi q)$ baseline is maximal at $p=0.19$, low $T$, and $\phi=0^\circ$, and it diminishes as $p$ is reduced or as either $T$ or $\phi$ increases - concomitant with the decrease of CDF spectral weight and of the phonon softening. Since the RIXS phonon intensity scales with the square of the EPC matrix element \cite{ament2011resonant, DevereauxPhon, BraicovichPhon}, these correlations further support a tight link in YBCO between CDF intensity, the magnitude of the BS-phonon softening, and EPC strength.

As an illustration of the strong link between CDF and phonon softening, Supplementary Fig. \ref{fig:IntBS} shows that only in the underdoped sample —either at $\phi=45^{\circ}$ or at $T=270$ K, where CDF contributions are minimized in our data — does the bond-stretching-phonon intensity follow the $\sin^{2}(\pi q)$ form within experimental uncertainty. This convergence reinforces the view that the observed intensity enhancement is driven primarily by coupling to dynamic CDF. Consequently, the joint temperature and angle dependent behaviour of the phonon intensity provides an independent confirmation that CDF dominate the modulation of the electron–phonon interaction, as discussed in the main text.


\subsection{\textbf{Volume of charge order peak}}\label{volumes}
Supplementary Fig.  \ref{fig:volumes} displays the temperature evolution of the integrated scattering volume associated with charge order - that is, the combined quasi -- elastic weight of CDF and CDW, when the latter is present -- for the two doping levels $p = 0.13$ and $p = 0.19$. The volume, $V$, is obtained from Lorentzian fits to the quasi-elastic peaks in momentum space (see Fig. 2(c),(d) in the main text) and is defined following Ref. \cite{ArpaiaUNO} as $V = A\,(\mathrm{FWHM})^{2}$, where $A$ is the peak amplitude and FWHM is its full width at half maximum, assuming the peak is isotropic along the $H$ and $K$ directions.
As highlighted in the main text, within experimental uncertainty, the extracted volumes are essentially independent of both temperature and doping. In particular, in the underdoped $p=0.13$ sample the total quasielastic volume (i) is not higher than in the fully oxygenated sample even at low temperature, where CDW is strongest; and (ii) is nearly temperature-independent, although at high temperature only weak CDF survive. This result indicates that -- even in the presence of the pronounced low-temperature CDW peak at $p = 0.13$ -- the quasielastic signal is dominated by the broader CDF component, whose integrated weight changes only marginally with either temperature or carrier concentration. It is therefore not surprising that the lattice renormalization is governed by the dynamic charge response rather than by quasi-static CDW.

\subsection{\textbf{Characteristic energy of CDF}}\label{energyCDF}
In Supplementary Fig. \ref{fig:energyCDF} the characteristic energy of CDF at $p = 0.19$ is presented as a function of temperature. This quantity was extracted using two independent approaches that yielded consistent results. The first, denoted as $\omega_0$, corresponds to the energy position of the Gaussian peak used to model the CDF contribution in the RIXS spectra (see also Fig. 4(b) in the main text). The second, $\omega_f$, was obtained indirectly from the quasi-elastic peak in momentum space,  by applying the relation $\omega_f \sim \nu_0 \cdot \mathrm{FWHM}^2$, with $\nu_0 = 1.26$~r.l.u., as reported in Ref.~\cite{NatCommRArpaia}. The excellent agreement between these two independent estimates confirms both the reliability of the extracted values and the robustness of the fitting procedure. 

Remarkably, the CDF energy increases with temperature, inversely tracking the evolution of $\Delta E$. As $\omega_{0}$ (or $\omega_f$) grows, the CDF peak approaches the bond-stretching-phonon branch, weakening their interference and thereby diminishing the phonon softening,  consistent with our observations.

\section{\label{sec: ack}Acknowledgments}

We are grateful for insightful discussions with Carlo Di Castro, Marco Grilli and Götz Seibold. The experimental data were collected at the beam line ID32 of the ESRF during experiments HC4149, HC5627 and HC6089. This work was performed in part at Myfab Chalmers. The work by G.M. was jointly supported by Politecnico di Milano and European X-ray Free Electron Laser Facility GmbH.
F.L. acknowledges support by the Swedish Research Council (VR) under the project 2022-04334. S.C. acknowledges support from the University of Rome Sapienza, projects Ateneo 2023 (RM123188E830D258), Ateneo 2024 (RM124190C54BE48D), and Ateneo 2025 (RP125199B9FDBFE4).

\section{\label{sec: author}Author Contributions}

R.A. and G.G. conceived and designed the experiments with suggestions from S.C. and F.L.. M.F., G.M., F.R., N.B.B., G.G. and R.A. performed the RIXS measurements. R.A. grew and characterized the YBCO films. M.F. and R.A. analysed the RIXS experimental data; M.F., R.A., S.C., G.G., M.M.S. and F.L. discussed and interpreted the results. M.F. and R.A. wrote the manuscript with  contributions from all authors.

\section{\label{sec: info}Additional Information}

Correspondence and requests for materials should be addressed to Martina Fedele, Giacomo Ghiringhelli and Riccardo Arpaia.  

\section{\label{sec:figures}Figures}
\begin{figure}[H]
    \centering
    \includegraphics[width=0.9\linewidth]{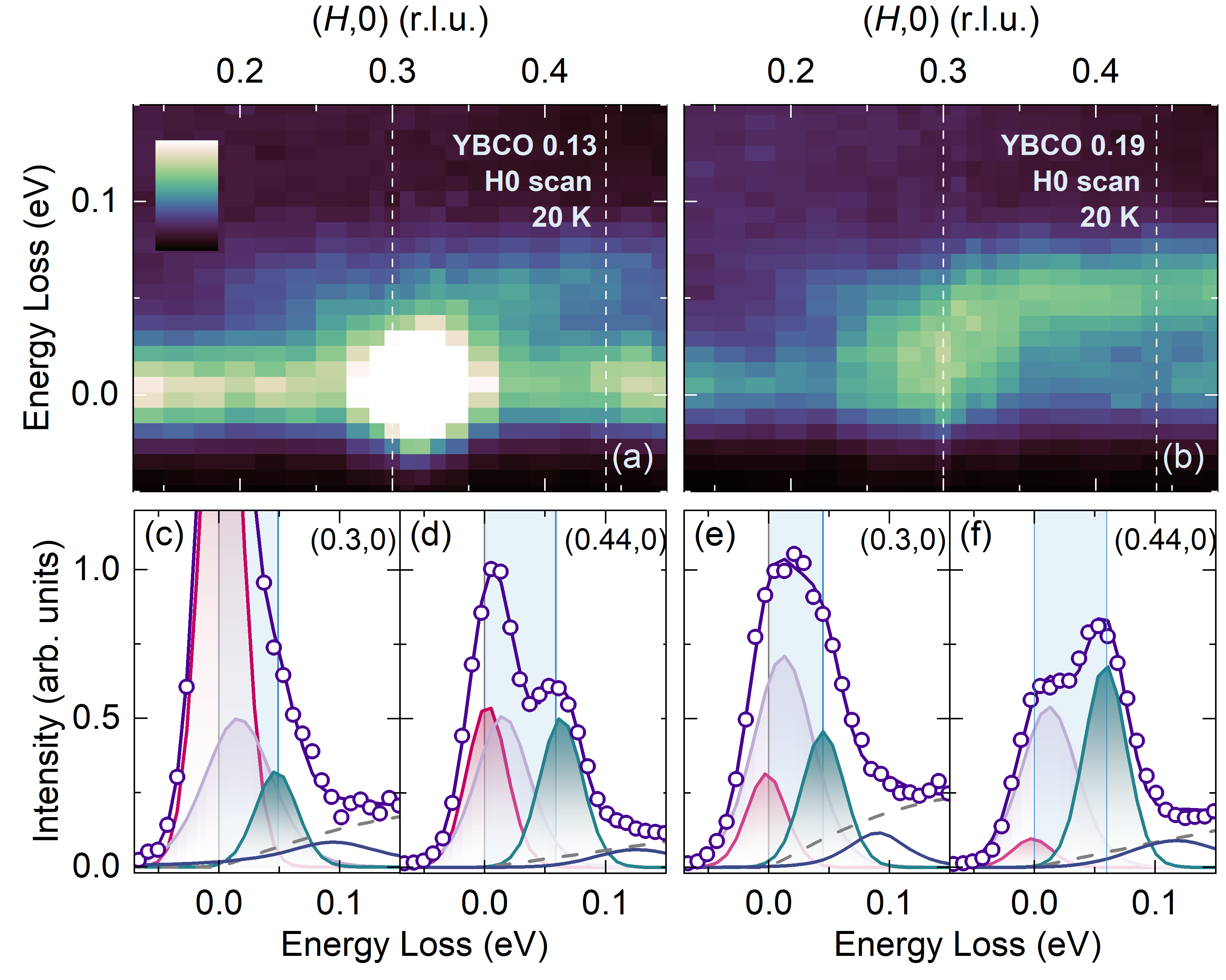}
    \caption{\textbf{Entwining of charge order and BS phonons measured by RIXS.} (a)-(b) Intensity maps acquired at 20~K on YBCO, at doping levels where charge order is dominated respectively by CDW ($p=0.13$) and CDF ($p=0.19$). (c)-(d) Fits of representative RIXS spectra for YBCO with $p = 0.13$ at two momentum values (indicated by dashed lines in the intensity map): $q = 0.30$, where the softening is most pronounced, and $q = 0.44$, where it is significantly reduced. 
    The change in energy of the BS phonons, represented by the green Gaussian, is highlighted by the light blue bar. The red, light violet, and blue Gaussians, as well as the region below the gray dashed line, correspond respectively to the elastic peak, the CDF contribution, the BS overtone, and the paramagnons. (e)-(f) Same as (c)-(d), but for YBCO $p=0.19$. 
    }   
    \label{fig:figure1}
\end{figure}
\begin{figure}[H]
    \centering
    \includegraphics[width=0.9\linewidth]{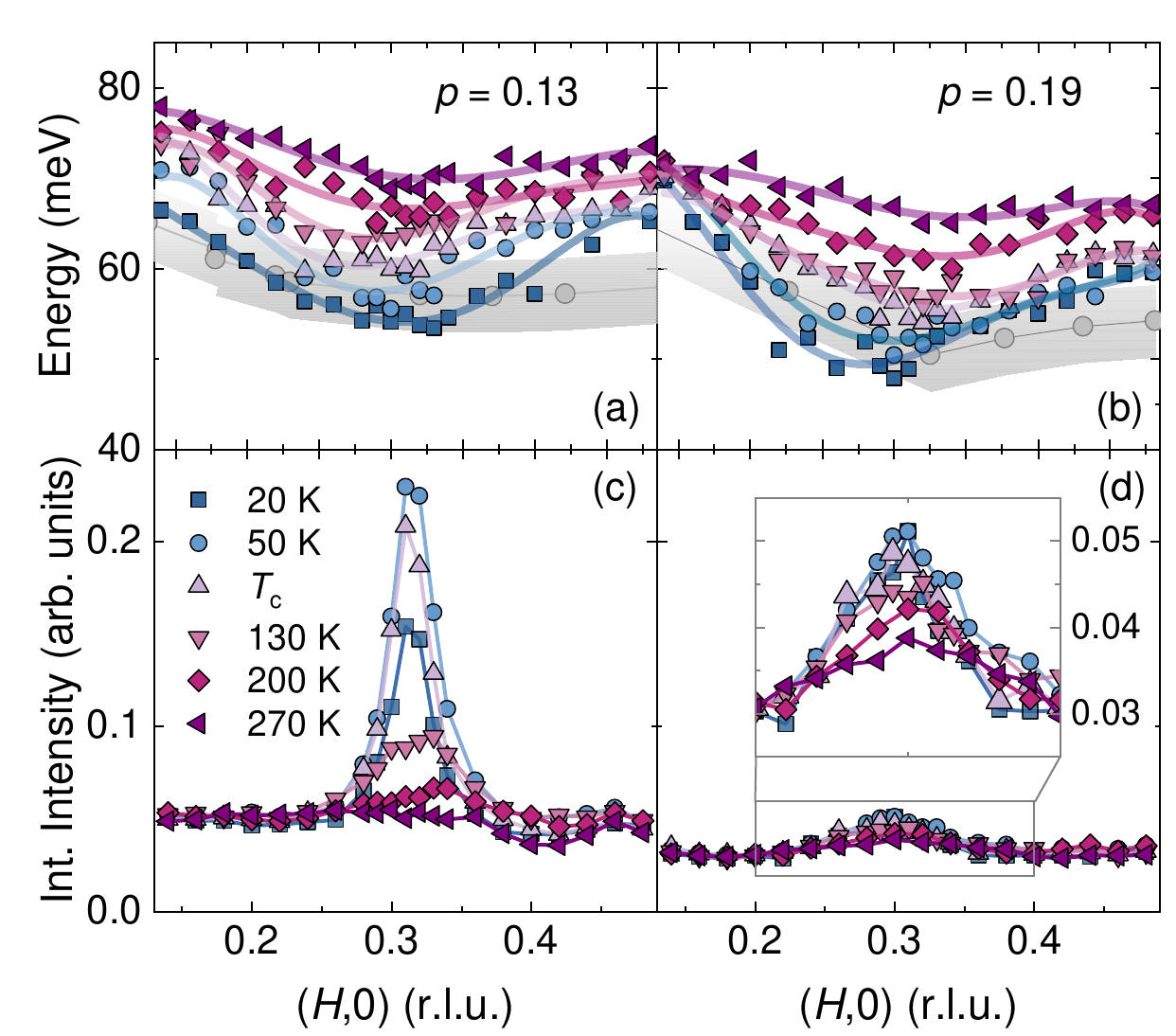}
    \caption{\textbf{Temperature evolution of charge order intensity and BS phonon energy.} (a)-(b) The energy of the BS phonons, determined from the Gaussian fits as in Fig.~\ref{fig:figure1}, plotted as a function of temperature and \textbf{q}, respectively for YBCO $p=0.13$ and $p=0.19$. Lines are guides to the eye. Previous INS measurements~\cite{Pintschovius2005}, performed at $T\approx10$~K, are overlaid in grey on our data for direct comparison. The grey shaded area indicates the 95\% confidence interval extracted from the fitting procedure.
    (c)-(d) Energy-integrated intensities of the quasielastic peak, evaluated in the range [–0.1, 0.035]~eV, respectively for YBCO $p=0.13$ and $p=0.19$. Only in the underdoped case the peak is dominated at low temperature (below 200~K) by CDW, while CDF persist at higher temperatures and at all temperatures in the fully oxygenated sample.}
    \label{fig:figure2}
\end{figure}

\begin{figure}[H]
    \centering
    \includegraphics[width=0.8\linewidth]{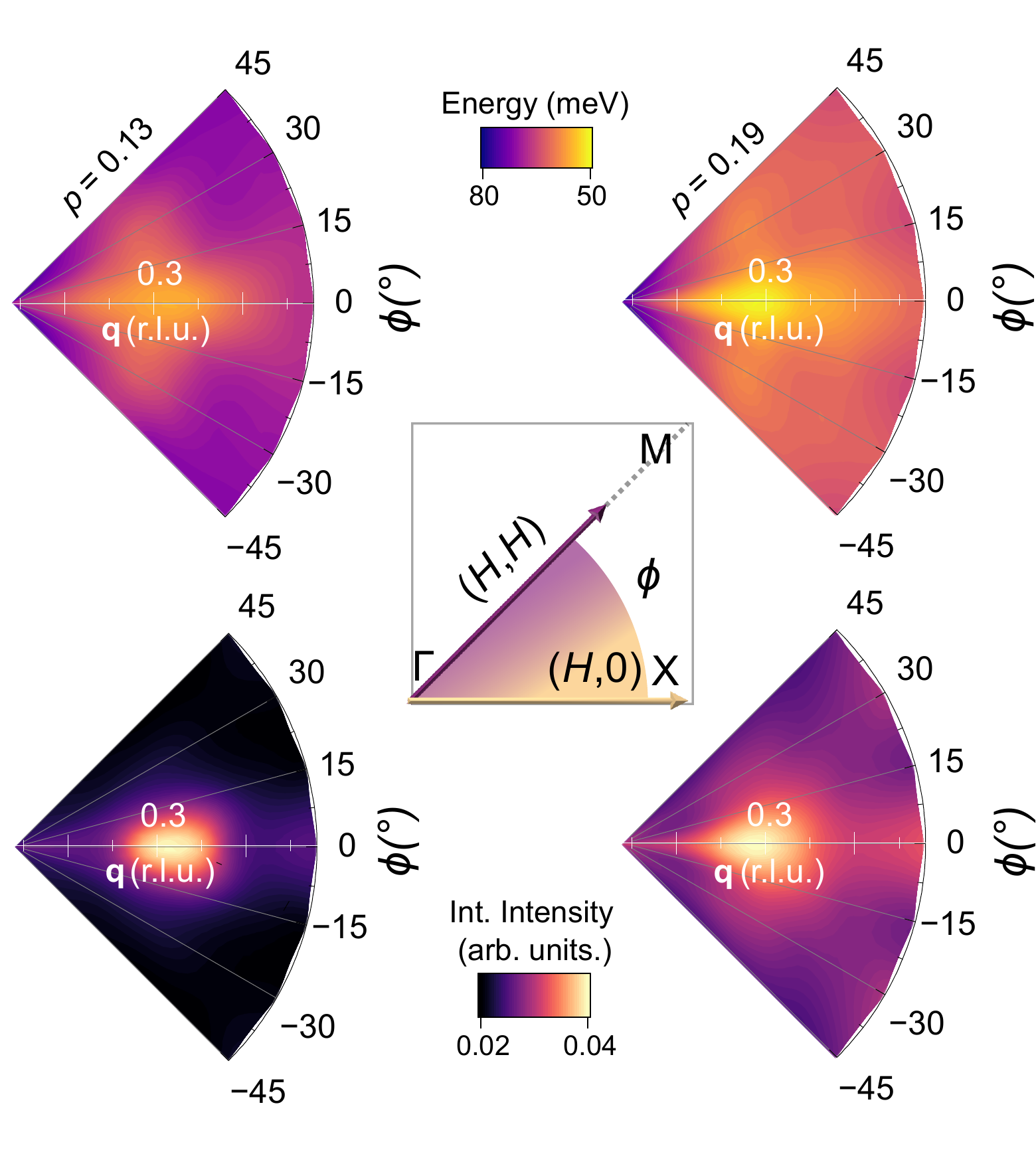}
    \caption{\textbf{Azimuthal evolution of charge order intensity and BS phonon energy.} (a)-(b) Maps for YBCO $p=0.13$ and $p=0.19$, showing the BS phonon energy as a function of momentum $q$ and azimuthal angle $\phi$, within the wedge-shaped region of the Brillouin zone connecting the $\Gamma$–X and $\Gamma$–M directions, as sketched in the central inset. (c)-(d) Maps for YBCO $p=0.13$ and $p=0.19$, showing the integrated intensity of the quasielastic peak in the energy range [–100, 35]~meV, across the same wedge-shaped region of the Brillouin zone as in (a)-(b).
    }
    \label{fig:figure3}
\end{figure}

\begin{figure}[H]
    \centering
    \includegraphics[width=0.9\linewidth]{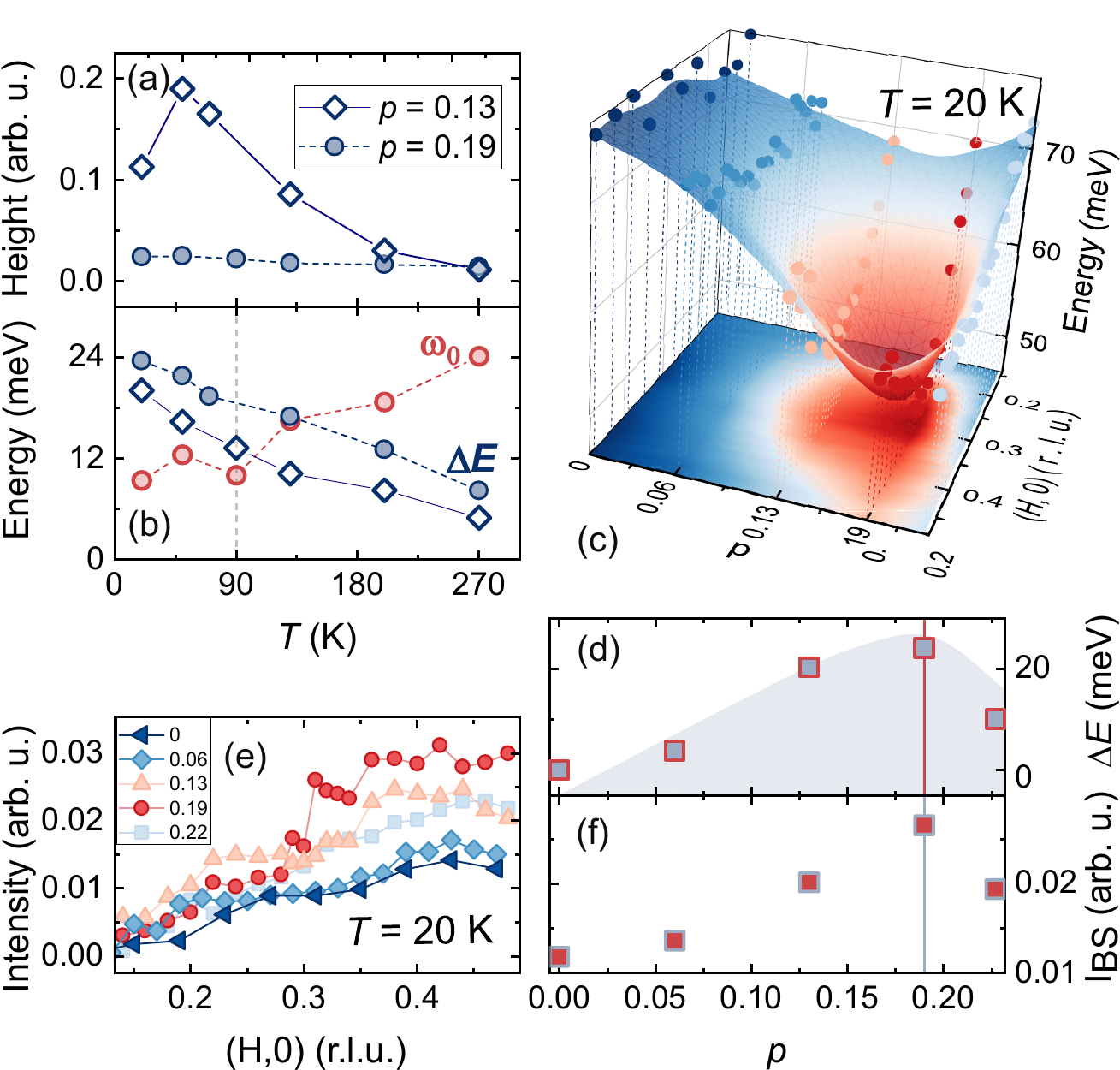}
    \caption{\textbf{BS phonon softening driven by CDF: converging fingerprints of superconductivity and electron–phonon coupling.} 
    (a) Height of the quasielastic peak and (b) magnitude of the BS phonon energy softening $\Delta E$ as a function of temperature for YBCO $p = 0.13$ (diamonds) and $p = 0.19$ (circles). In panel (b), the temperature dependence of the CDF energy $\omega_0$ for $p = 0.19$ is also shown (red circles).
    The vertical dashed line indicates $T_\mathrm{c}$ at $p = 0.19$. 
    (c) BS phonon energy as a function of momentum transfer $q$ and doping $p$, measured at $T = 20$~K. The 3D surface highlights a pronounced softening at $p = 0.19$, where both the CDF intensity and the superconducting strength reach their maximum. Experimental data points are overlaid on the surface.
    (d) Doping dependence of $\Delta E$ at $T=20$~K. The data closely follow the doping dependence of the CDF peak height, as reported in Ref.~\cite{NatCommRArpaia} (light blue region). (e) Momentum dependence of the BS phonon intensity for different doping levels at $T = 20$~K. The average intensity over a broad momentum range ($q > 0.3$~r.l.u.) defines the integrated quantity $I_{\mathrm{BS}}$. (f) Doping dependence of $I_{\mathrm{BS}}$ at $T = 20$~K, which mirrors the behavior of both the CDF peak height and the phonon softening $\Delta E$.}
    \label{fig:figure4}
\end{figure}

\section{Supplementary Figures}

\newenvironment{myextfigure}[1][]%
  {\captionsetup{type=extfigure}\begin{figure}[#1]}%
  {\end{figure}}

\begin{extfigure}[H]
    \centering
    \includegraphics[width=0.5\linewidth]{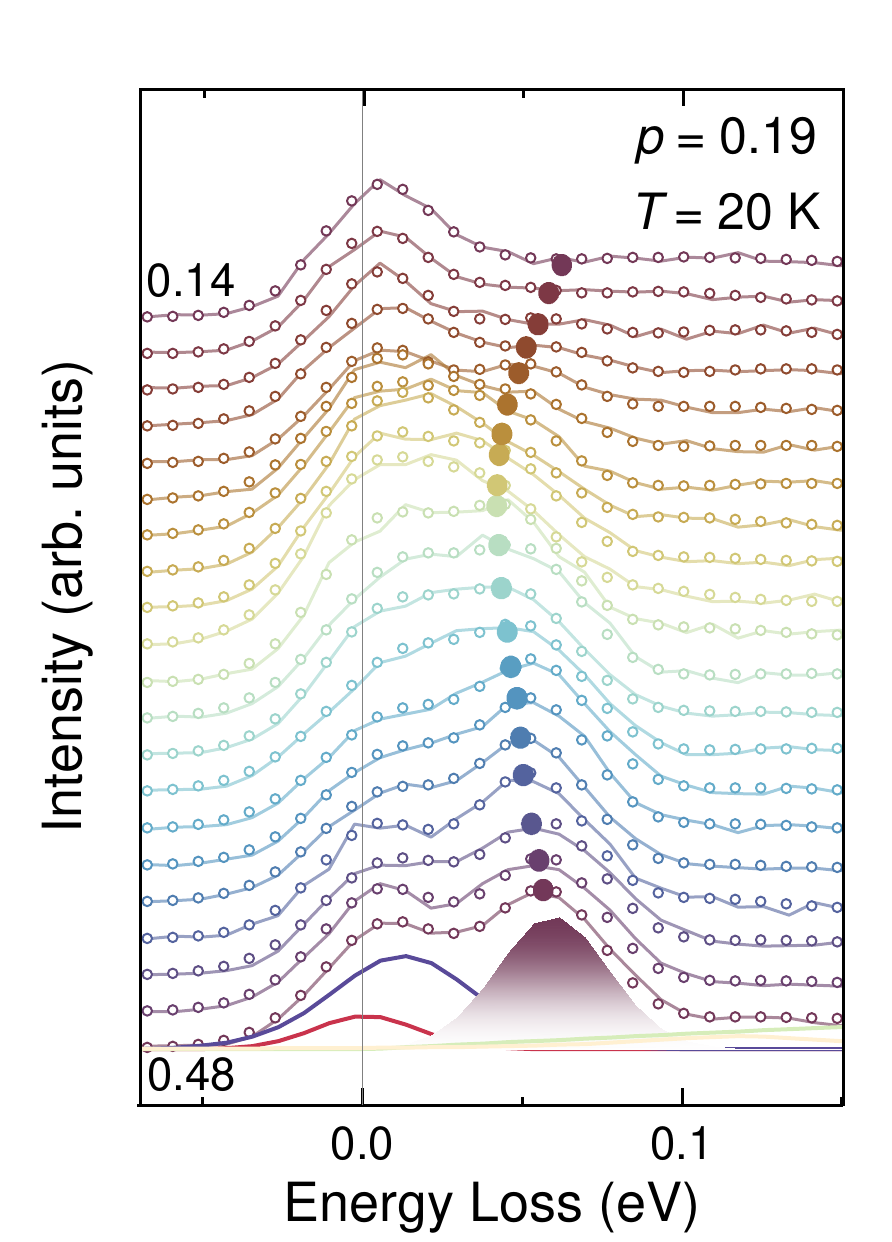}
    \caption{Multi-peak fitting of the $T = 20$~K RIXS spectra for the slightly overdoped YBCO sample at $p = 0.19$. Experimental data are shown as emptly circles, while the corresponding fit curves are represented by solid lines. For each value of transferred momentum $q$ in the range $[0.14, 0.48]$~r.l.u., the energy position of the violet Gaussian peak associated with the bond-stretching (BS) phonon is indicated by filled circles. A clear softening of the phonon mode is observed around $q \sim 0.3$~r.l.u.
}
    \label{fig:fitting}
\end{extfigure}

\begin{extfigure}[H]
    \centering
    \includegraphics[width=0.6\linewidth]{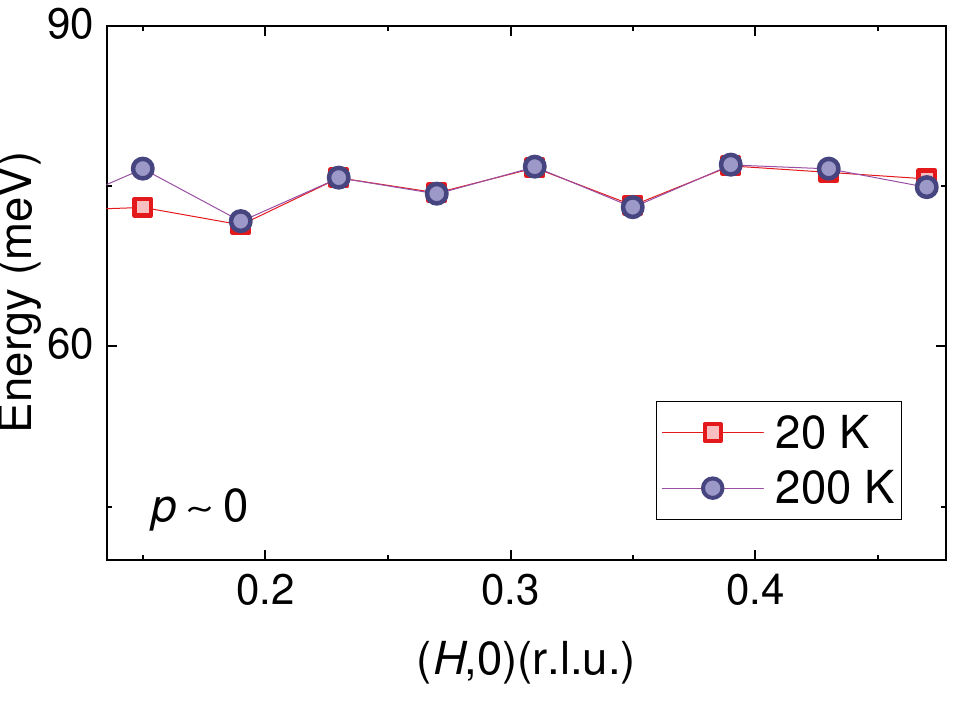}
    \caption{Temperature dependence of the phonon dispersion in the undoped sample ($p\approx0$). Within the experimental uncertainty, no discernible energy renormalization is observed over the entire temperature range explored.
  }
    \label{fig:Dopdep}
\end{extfigure}
\begin{extfigure}[H]
    \centering
    \includegraphics[width=0.4\linewidth]{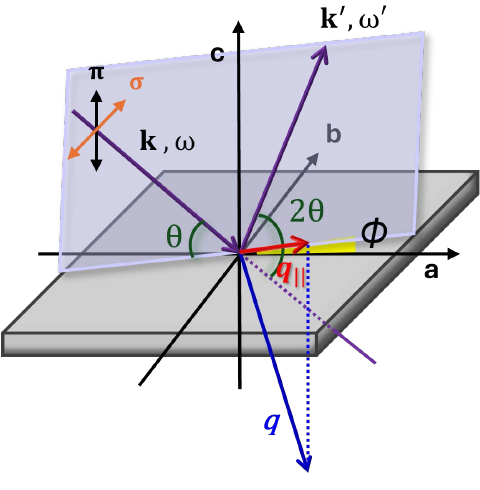}
    \caption{RIXS scattering geometry.}
    \label{fig:RIXSgeometry}
\end{extfigure}

\begin{extfigure}[H]
    \centering
    \includegraphics[width=0.8\linewidth]{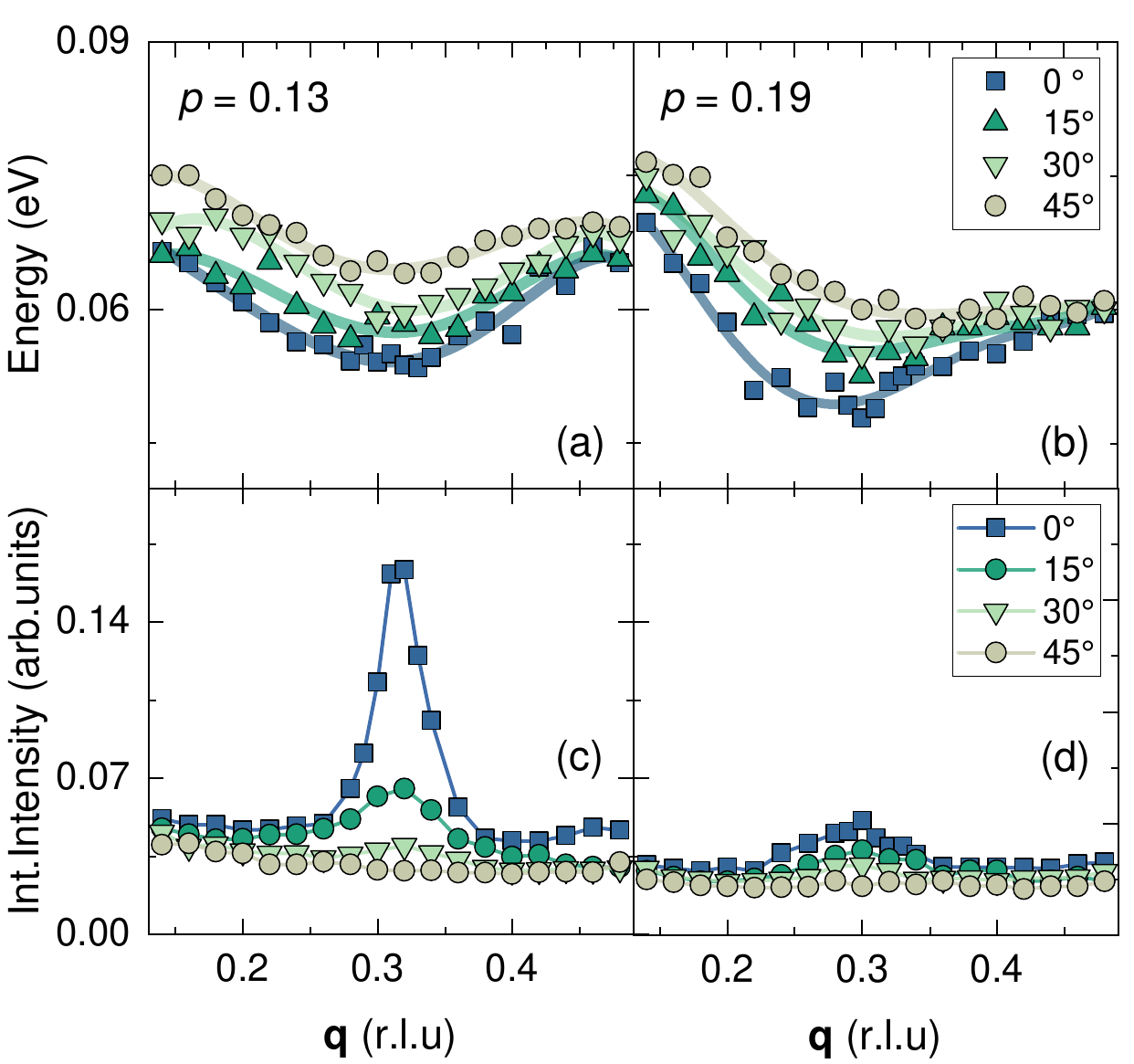}
    \caption{Azimuthal‐angle dependence at $T = 20$ K. Panels (a) and (b) display the bond-stretching‐phonon dispersion for the doping levels $p = 0.13$ and $p = 0.19$, respectively, whereas panels (c) and (d) show the associated quasielastic integrated intensities.
}
    \label{fig:Phiscan}
\end{extfigure}
\begin{extfigure}[H]
    \centering
    \includegraphics[width=0.8\linewidth]{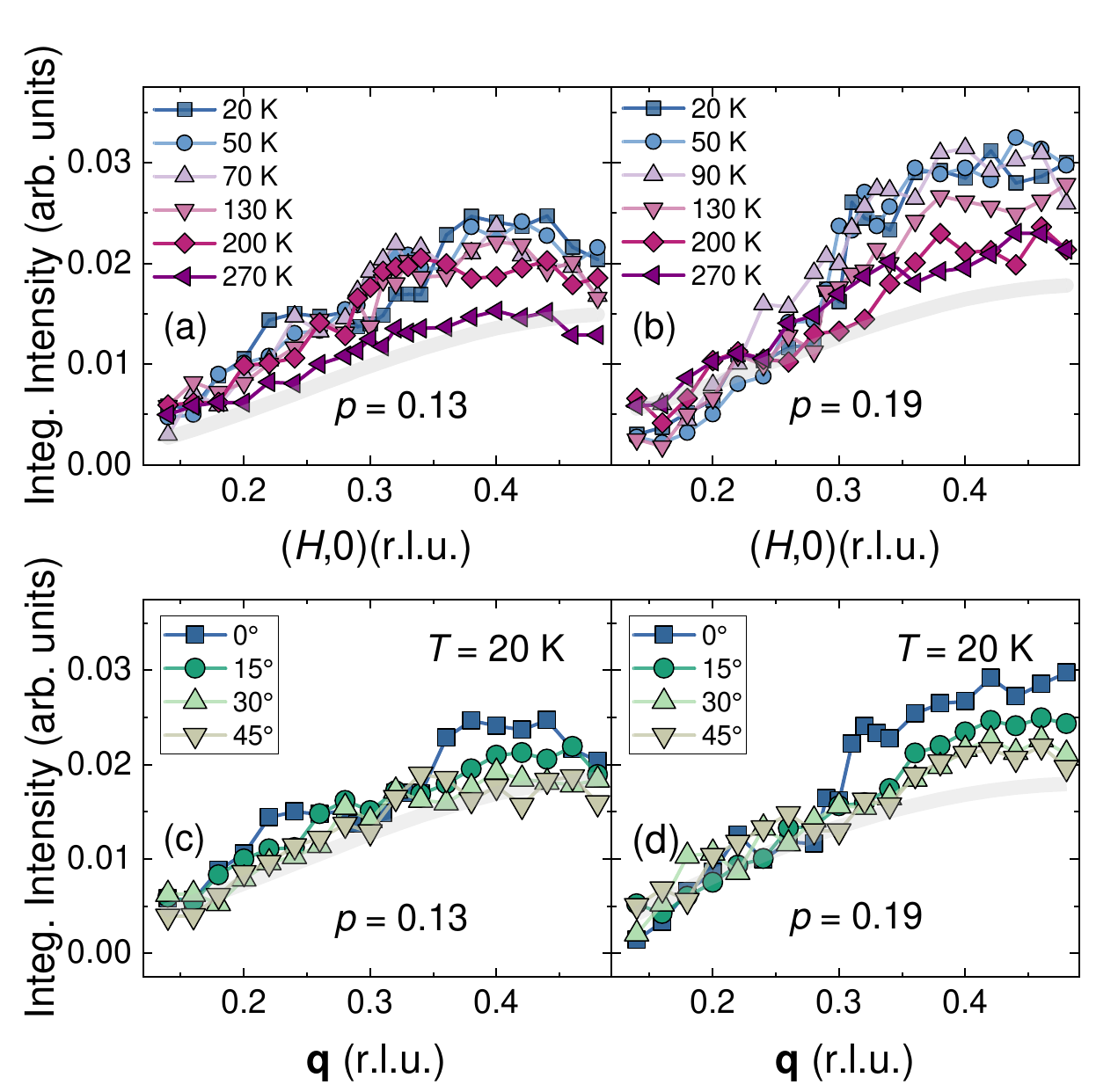}
    \caption{Integrated intensity of the Gaussian peak associated to bond stretching phonon. An intensity anomaly appears relative to the expected behavior proportional to $\sin^2(\pi q)$, observed along the $(H,H)$ direction (dashed gray line). As a consequence of the phonon softening this anomaly shows both temperature (a)-(b) and azimuthal angle (c)-(d) dependence. 
}
    \label{fig:IntBS}
\end{extfigure}
\begin{extfigure}[H]
    \centering
    \includegraphics[width=0.61\linewidth]{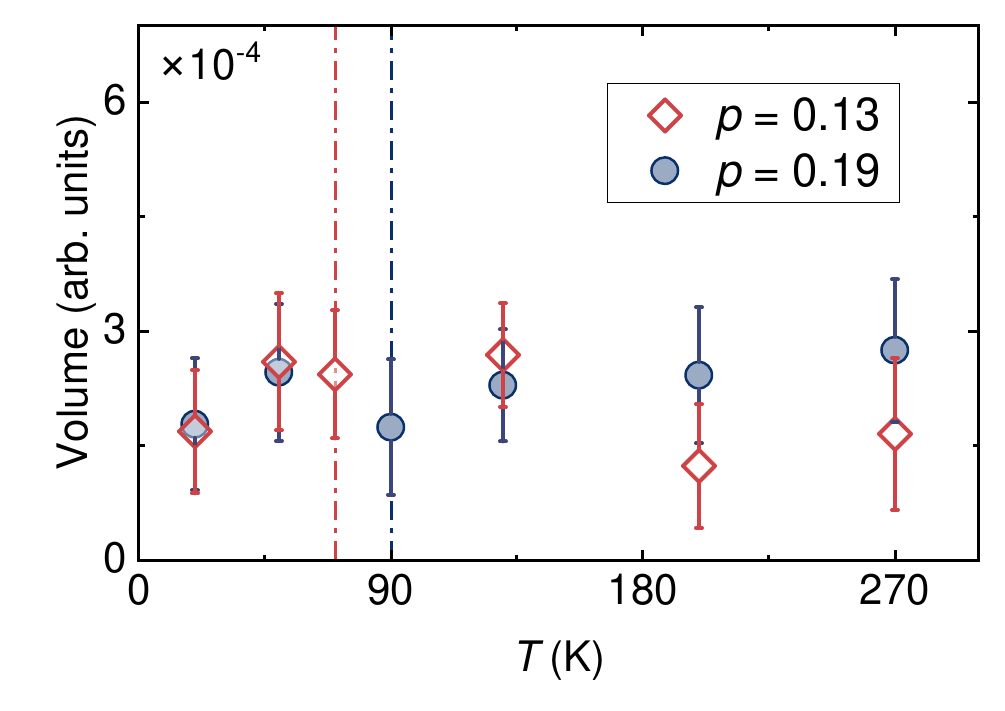}
    \caption{Temperature evolution of the quasielastic-peak volume, obtained by integrating the scattering intensity over the energy range [–35, 100]~meV. This volume represents the total charge-order spectral weight (CDF + CDW). Although the CDW signal reaches its maximum at $p = 0.13$, its contribution to the integrated volume is negligible: the total quasielastic weight (i.e., the volume) is thus dominated by the broader CDF component.
}
    \label{fig:volumes}
\end{extfigure}
\begin{extfigure}[H]
    \centering
    \includegraphics[width=0.6\linewidth]{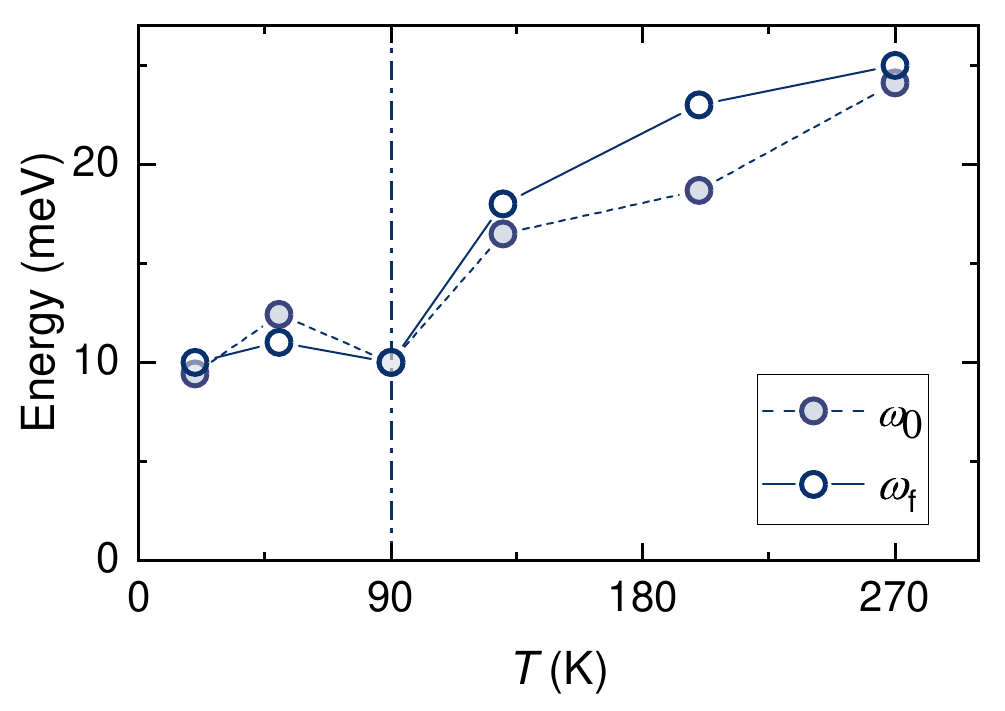}
    \caption{Temperature dependence of the CDF energy at $q = q_{\mathrm{CDF}}$ in YBCO with $p = 0.19$. The values are extracted directly from the fits, using the position of the Gaussian peak associated with CDFs ($\omega_0$, full symbols), and indirectly from the FWHM of the quasielastic peak ($\omega_f$, open symbols), by applying the relation $\omega_f \sim \nu_0 \cdot \mathrm{FWHM}^2$ with $\nu_0 = 1.26$~r.l.u., as reported in Ref.~\cite{NatCommRArpaia}. The two determinations, in very good qualitative agreement, both show a monotonic increase of the CDF energy with temperature. The vertical dashed line marks the superconducting critical temperature $T_c$.}

    \label{fig:energyCDF}
\end{extfigure}

\bibliographystyle{unsrtnat} 
\bibliography{BS_YBCO_ID32}

\begin{thebibliography}{78}
\providecommand{\natexlab}[1]{#1}
\providecommand{\url}[1]{\texttt{#1}}
\expandafter\ifx\csname urlstyle\endcsname\relax
  \providecommand{\doi}[1]{doi: #1}\else
  \providecommand{\doi}{doi: \begingroup \urlstyle{rm}\Url}\fi

\bibitem[cas(1996)]{castellani1996non}
Non-fermi-liquid behavior and d-wave superconductivity near the charge-density-wave quantum critical point.
\newblock \emph{Zeitschrift f{\"u}r Physik B Condensed Matter}, 103:\penalty0 137--144, 1996.

\bibitem[Neto(2001)]{neto2001charge}
AH~Castro Neto.
\newblock Charge density wave, superconductivity, and anomalous metallic behavior in 2d transition metal dichalcogenides.
\newblock \emph{Phys. Rev. Lett.}, 86\penalty0 (19):\penalty0 4382, 2001.
\newblock URL \url{https://doi.org/10.1103/PhysRevLett.86.4382}.

\bibitem[Fradkin et~al.(2015)Fradkin, Kivelson, and Tranquada]{fradkin2015colloquium}
Eduardo Fradkin, Steven~A Kivelson, and John~M Tranquada.
\newblock Colloquium: Theory of intertwined orders in high temperature superconductors.
\newblock \emph{Rev. Mod. Phys.}, 87\penalty0 (2):\penalty0 457--482, 2015.
\newblock URL \url{https://doi.org/10.1103/RevModPhys.87.457}.

\bibitem[Wang et~al.(2022)Wang, He, Yang, Garcia-Fernandez, Nag, Zhou, Minola, Tacon, Keimer, Peng, et~al.]{wang2022paramagnons}
Lichen Wang, Guanhong He, Zichen Yang, Mirian Garcia-Fernandez, Abhishek Nag, Kejin Zhou, Matteo Minola, Matthieu~Le Tacon, Bernhard Keimer, Yingying Peng, et~al.
\newblock Paramagnons and high-temperature superconductivity in a model family of cuprates.
\newblock \emph{Nat. Commun.}, 13\penalty0 (1):\penalty0 3163, 2022.
\newblock URL \url{https://doi.org/10.1038/s41467-022-30918-z}.

\bibitem[Neupert et~al.(2022)Neupert, Denner, Yin, Thomale, and Hasan]{neupert2022charge}
Titus Neupert, M~Michael Denner, Jia-Xin Yin, Ronny Thomale, and M~Zahid Hasan.
\newblock Charge order and superconductivity in kagome materials.
\newblock \emph{Nat. Phys.}, 18\penalty0 (2):\penalty0 137--143, 2022.
\newblock URL \url{https://doi.org/10.1038/s41567-021-01404-y}.

\bibitem[Keimer et~al.(2015)Keimer, Kivelson, Norman, Uchida, and Zaanen]{keimer2015quantum}
Bernhard Keimer, Steven~A Kivelson, Michael~R Norman, Shinichi Uchida, and J~Zaanen.
\newblock {From quantum matter to high-temperature superconductivity in copper oxides}.
\newblock \emph{Nature}, 518\penalty0 (7538):\penalty0 179--186, 2015.
\newblock URL \url{https://doi.org/10.1038/nature14165}.

\bibitem[Lanzara et~al.(2001)Lanzara, Bogdanov, Zhou, Kellar, Feng, Lu, Yoshida, Eisaki, Fujimori, Kishio, et~al.]{lanzara2001evidence}
A~Lanzara, PV~Bogdanov, XJ~Zhou, SA~Kellar, DL~Feng, ED~Lu, Teppei Yoshida, H~Eisaki, Atsushi Fujimori, Kohji Kishio, et~al.
\newblock Evidence for ubiquitous strong electron--phonon coupling in high-temperature superconductors.
\newblock \emph{Nature}, 412\penalty0 (6846):\penalty0 510--514, 2001.
\newblock URL \url{https://doi.org/10.1038/35087518}.

\bibitem[Kim et~al.(2002)Kim, Ronning, Damascelli, Feng, Shen, Wells, Kim, Birgeneau, Kastner, Miller, et~al.]{kim2002anomalous}
C~Kim, F~Ronning, A~Damascelli, DL~Feng, Z-X Shen, BO~Wells, YJ~Kim, RJ~Birgeneau, MA~Kastner, LL~Miller, et~al.
\newblock Anomalous temperature dependence in the photoemission spectral function of cuprates.
\newblock \emph{Phys. Rev. B}, 65\penalty0 (17):\penalty0 174516, 2002.
\newblock URL \url{https://doi.org/10.1103/PhysRevB.65.174516}.

\bibitem[Shen et~al.(2004)Shen, Ronning, Lu, Lee, Ingle, Meevasana, Baumberger, Damascelli, Armitage, Miller, et~al.]{shen2004missing}
KM~Shen, F~Ronning, DH~Lu, WS~Lee, NJC Ingle, W~Meevasana, F~Baumberger, A~Damascelli, NP~Armitage, LL~Miller, et~al.
\newblock Missing quasiparticles and the chemical potential puzzle in the doping evolution of the cuprate superconductors.
\newblock \emph{Phys. Rev. Lett.}, 93\penalty0 (26):\penalty0 267002, 2004.
\newblock URL \url{https://doi.org/10.1103/PhysRevLett.93.267002}.

\bibitem[Ronning et~al.(2005)Ronning, Shen, Armitage, Damascelli, Lu, Shen, Miller, and Kim]{ronning2005anomalous}
F~Ronning, KM~Shen, NP~Armitage, A~Damascelli, DH~Lu, Z-X Shen, LL~Miller, and C~Kim.
\newblock {Anomalous high-energy dispersion in angle-resolved photoemission spectra from the insulating cuprate Ca$_2$CuO$_2$Cl$_2$}.
\newblock \emph{Phys. Rev. B}, 71\penalty0 (9):\penalty0 094518, 2005.
\newblock URL \url{https://doi.org/10.1103/PhysRevB.71.094518}.

\bibitem[Kogar et~al.(2017)Kogar, de~La~Pena, Lee, Fang, Sun, Lioi, Karapetrov, Finkelstein, Ruff, Abbamonte, et~al.]{kogar2017observation}
Anshul Kogar, Gilberto~Antonio de~La~Pena, Sangjun Lee, Yizhi Fang, SX-L Sun, David~B Lioi, Goran Karapetrov, Kenneth~D Finkelstein, Jacob~PC Ruff, Peter Abbamonte, et~al.
\newblock {Observation of a charge density wave incommensuration near the superconducting dome in Cu$_x$TiSe$_2$}.
\newblock \emph{Phys. Rev. Lett.}, 118\penalty0 (2):\penalty0 027002, 2017.
\newblock URL \url{https://doi.org/10.1103/PhysRevLett.118.027002}.

\bibitem[Hinlopen et~al.(2024)Hinlopen, Moulding, Broad, Buhot, Bangma, McCollam, Ayres, Sayers, Da~Como, Flicker, et~al.]{hinlopen2024lifshitz}
Roemer~DH Hinlopen, Owen~N Moulding, William~R Broad, Jonathan Buhot, Femke Bangma, Alix McCollam, Jake Ayres, Charles~J Sayers, Enrico Da~Como, Felix Flicker, et~al.
\newblock Lifshitz transition enabling superconducting dome around a charge-order critical point.
\newblock \emph{Sci. Adv.}, 10\penalty0 (27):\penalty0 eadl3921, 2024.
\newblock URL \url{https://www.science.org/doi/full/10.1126/sciadv.adl3921}.

\bibitem[Tian and Savrasov(2025)]{PhysRevB.CDFKagome}
Yuan Tian and Sergey~Y. Savrasov.
\newblock {Unconventional superconductivity via charge fluctuations in the kagome metal CsV$_{3}$Sb$_{5}$}.
\newblock \emph{Phys. Rev. B}, 111:\penalty0 134507, Apr 2025.
\newblock \doi{10.1103/PhysRevB.111.134507}.
\newblock URL \url{https://link.aps.org/doi/10.1103/PhysRevB.111.134507}.

\bibitem[Kongruengkit et~al.(2025)Kongruengkit, Salinas, Pokharel, Ortiz, Wilson, and Harter]{kongruengkit2025kagomeCDF}
Terawit Kongruengkit, Andrea N.~Capa Salinas, Ganesh Pokharel, Brenden~R. Ortiz, Stephen~D. Wilson, and John~W. Harter.
\newblock Persistence of charge density wave fluctuations in the absence of long-range order in a hole-doped kagome metal, 2025.
\newblock URL \url{https://arxiv.org/abs/2508.13290}.

\bibitem[Chaix et~al.(2017)Chaix, Ghiringhelli, Peng, Hashimoto, Moritz, Kummer, Brookes, He, Chen, Ishida, et~al.]{chaix2017dispersive}
Laura Chaix, Giacomo Ghiringhelli, YY~Peng, Makoto Hashimoto, Brian Moritz, Kurt Kummer, Nick~Ben Brookes, Yu~He, Sudi Chen, Shigeyuki Ishida, et~al.
\newblock {Dispersive charge density wave excitations in Bi$_2$Sr$_2$CaCu$_2$O$_{8+\delta}$}.
\newblock \emph{Nat. Phys.}, 13\penalty0 (10):\penalty0 952--956, 2017.
\newblock URL \url{https://doi.org/10.1038/nphys4157}.

\bibitem[Li et~al.(2020)Li, Nag, Pelliciari, Robarts, Walters, Garcia-Fernandez, Eisaki, Song, Ding, Johnston, et~al.]{li2020multiorbital}
Jiemin Li, Abhishek Nag, Jonathan Pelliciari, Hannah Robarts, Andrew Walters, Mirian Garcia-Fernandez, Hiroshi Eisaki, Dongjoon Song, Hong Ding, Steven Johnston, et~al.
\newblock {Multiorbital charge-density wave excitations and concomitant phonon anomalies in Bi$_2$Sr$_2$LaCuO$_{6+ \delta}$}.
\newblock \emph{Proc. Natl. Acad. Sci. U.S.A.}, 117\penalty0 (28):\penalty0 16219--16225, 2020.
\newblock URL \url{https://doi.org/10.1073/pnas.2001755117}.

\bibitem[Lin et~al.(2020)Lin, Miao, Mazzone, Gu, Nag, Walters, Garc{\'\i}a-Fern{\'a}ndez, Barbour, Pelliciari, Jarrige, et~al.]{lin2020strongly}
JQ~Lin, H~Miao, DG~Mazzone, GD~Gu, A~Nag, AC~Walters, M~Garc{\'\i}a-Fern{\'a}ndez, A~Barbour, J~Pelliciari, I~Jarrige, et~al.
\newblock {Strongly correlated charge density wave in La$_{2-x}$Sr$_x$CuO$_4$ evidenced by doping-dependent phonon anomaly}.
\newblock \emph{Phys. Rev. Lett.}, 124\penalty0 (20):\penalty0 207005, 2020.
\newblock URL \url{https://doi.org/10.1103/PhysRevLett.124.207005}.

\bibitem[Huang et~al.(2021)Huang, Singh, Mou, Johnston, Kemper, van~den Brink, Chen, Lee, Okamoto, Chu, Li, Komiya, Komarek, Fujimori, Chen, and Huang]{Huang2021}
H.Y. Huang, A.~Singh, C.Y. Mou, S.~Johnston, A.F. Kemper, J.~van~den Brink, P.J. Chen, T.K. Lee, J.~Okamoto, Y.Y. Chu, J.H. Li, S.~Komiya, A.C. Komarek, A.~Fujimori, C.T. Chen, and D.J. Huang.
\newblock Quantum fluctuations of charge order induce phonon softening in a superconducting cuprate.
\newblock \emph{Phys. Rev. X}, 11\penalty0 (4):\penalty0 041038, 2021.
\newblock \doi{10.1103/PhysRevX.11.041038}.
\newblock URL \url{https://doi.org/10.1103/PhysRevX.11.041038}.

\bibitem[Wang et~al.(2021)Wang, von Arx, Horio, Mukkattukavil, K{\"u}spert, Sassa, Schmitt, Nag, Pyon, Takayama, et~al.]{wang2021charge}
Qisi Wang, Karin von Arx, Masafumi Horio, Deepak~John Mukkattukavil, Julia K{\"u}spert, Yasmine Sassa, Thorsten Schmitt, Abhishek Nag, Sunseng Pyon, Tomohiro Takayama, et~al.
\newblock {Charge order lock-in by electron-phonon coupling in La$_{1.675}$Eu$_{0.2}$Sr$_{0.125}$CuO$_4$}.
\newblock \emph{Sci. Adv.}, 7\penalty0 (27):\penalty0 eabg7394, 2021.
\newblock \doi{10.1126/sciadv.abg7394}.
\newblock URL \url{https://www.science.org/doi/full/10.1126/sciadv.abg7394}.

\bibitem[Lee et~al.(2021)Lee, Zhou, Hepting, Li, Nag, Walters, Garcia-Fernandez, Robarts, Hashimoto, Lu, et~al.]{lee2021spectroscopic}
Wei-Sheng Lee, Ke-Jin Zhou, M~Hepting, J~Li, A~Nag, AC~Walters, M~Garcia-Fernandez, HC~Robarts, M~Hashimoto, H~Lu, et~al.
\newblock {Spectroscopic fingerprint of charge order melting driven by quantum fluctuations in a cuprate}.
\newblock \emph{Nat. Phys.}, 17\penalty0 (1):\penalty0 53--57, 2021.
\newblock URL \url{https://doi.org/10.1038/s41567-020-0993-7}.

\bibitem[Lu et~al.(2022)Lu, Hashimoto, Chen, Ishida, Song, Eisaki, Nag, Garcia-Fernandez, Arpaia, Ghiringhelli, Braicovich, Zaanen, Moritz, Kummer, Brookes, Zhou, Shen, Devereaux, and Lee]{WSL_PRB_softenining}
Haiyu Lu, Makoto Hashimoto, Su-Di Chen, Shigeyuki Ishida, Dongjoon Song, Hiroshi Eisaki, Abhishek Nag, Mirian Garcia-Fernandez, Riccardo Arpaia, Giacomo Ghiringhelli, Lucio Braicovich, Jan Zaanen, Brian Moritz, Kurt Kummer, Nicholas~B. Brookes, Ke-Jin Zhou, Zhi-Xun Shen, Thomas~P. Devereaux, and Wei-Sheng Lee.
\newblock Identification of a characteristic doping for charge order phenomena in {B}i-2212 cuprates via rixs.
\newblock \emph{Phys. Rev. B}, 106:\penalty0 155109, Oct 2022.
\newblock \doi{10.1103/PhysRevB.106.155109}.
\newblock URL \url{https://link.aps.org/doi/10.1103/PhysRevB.106.155109}.

\bibitem[Tam et~al.(2022)Tam, Zhu, Ayres, Kummer, Yakhou-Harris, Cooper, Carrington, and Hayden]{tam2022charge}
CC~Tam, Mengze Zhu, Jake Ayres, Kurt Kummer, Flora Yakhou-Harris, JR~Cooper, Antony Carrington, and SM~Hayden.
\newblock {Charge density waves and Fermi surface reconstruction in the clean overdoped cuprate superconductor Tl$_2$Ba$_2$CuO$_{6+\delta}$}.
\newblock \emph{Nat. Commun.}, 13\penalty0 (1):\penalty0 570, 2022.
\newblock URL \url{https://doi.org/10.1038/s41467-022-28124-y}.

\bibitem[Scott et~al.(2023)Scott, Kisiel, Boyle, Basak, Jargot, Das, Agrestini, Garcia-Fernandez, Choi, Pelliciari, et~al.]{scott2023low}
Kirsty Scott, Elliot Kisiel, Timothy~J Boyle, Rourav Basak, Ga{\"e}tan Jargot, Sarmistha Das, Stefano Agrestini, Mirian Garcia-Fernandez, Jaewon Choi, Jonathan Pelliciari, et~al.
\newblock {Low-energy quasi-circular electron correlations with charge order wavelength in Bi$_2$Sr$_2$CaCu$_2$O$_{8+\delta}$}.
\newblock \emph{Sci. Adv.}, 9\penalty0 (29):\penalty0 eadg3710, 2023.
\newblock URL \url{https://www.science.org/doi/full/10.1126/sciadv.adg3710}.

\bibitem[Pintschovius et~al.(1991)Pintschovius, Pyka, Reichardt, Rumiantsev, Mitrofanov, Ivanov, Collin, and Bourges]{PintschoviusTRE}
L.~Pintschovius, N.~Pyka, W.~Reichardt, A.Yu. Rumiantsev, N.L. Mitrofanov, A.S. Ivanov, G.~Collin, and P.~Bourges.
\newblock {Lattice dynamical studies of HTSC materials}.
\newblock \emph{Physica C}, 185-189:\penalty0 156--161, 1991.
\newblock ISSN 0921-4534.
\newblock \doi{https://doi.org/10.1016/0921-4534(91)91965-7}.
\newblock URL \url{https://www.sciencedirect.com/science/article/pii/0921453491919657}.

\bibitem[Reichardt et~al.(1994)Reichardt, Pintschovius, Pyka, Schweiss, Erb, Bourges, Collin, Rossat-Mignod, Henry, Ivanov, et~al.]{reichardt1994anharmonicity}
W~Reichardt, L~Pintschovius, N~Pyka, P~Schweiss, A~Erb, P~Bourges, G~Collin, J~Rossat-Mignod, IY~Henry, AS~Ivanov, et~al.
\newblock Anharmonicity and electron-phonon coupling in cuprate superconductors studied by inelastic neutron scattering.
\newblock \emph{J. Supercond.}, 7\penalty0 (2):\penalty0 399--407, 1994.
\newblock URL \url{https://link.springer.com/article/10.1007/BF00724577}.

\bibitem[Chaplot et~al.(1995)Chaplot, Reichardt, Pintschovius, and Pyka]{Chaplot1995}
S.~L. Chaplot, W.~Reichardt, L.~Pintschovius, and N.~Pyka.
\newblock Common interatomic potential model for the lattice dynamics of several cuprates.
\newblock \emph{Phys. Rev. B}, 52\penalty0 (10):\penalty0 7230--7242, September 1995.
\newblock \doi{10.1103/PhysRevB.52.7230}.
\newblock URL \url{https://doi.org/10.1103/PhysRevB.52.7230}.

\bibitem[Pintschovius et~al.(2002)Pintschovius, Reichardt, Kläser, Wolf, and v.~Löhneysen]{Pintschovius2002}
L.~Pintschovius, W.~Reichardt, M.~Kläser, T.~Wolf, and H.~v.~Löhneysen.
\newblock {Pronounced In-Plane Anisotropy of Phonon Anomalies in YBa\textsubscript{2}Cu\textsubscript{3}O\textsubscript{6.6}}.
\newblock \emph{Phys. Rev. Lett.}, 89\penalty0 (3):\penalty0 037001, July 2002.
\newblock \doi{10.1103/PhysRevLett.89.037001}.
\newblock URL \url{https://doi.org/10.1103/PhysRevLett.89.037001}.

\bibitem[Pintschovius(2003)]{Pintschovius2003}
Lothar Pintschovius.
\newblock {Search for Signature of Charge Inhomogeneities in the Phonons of YBa\textsubscript{2}Cu\textsubscript{3}O\textsubscript{7--x}}.
\newblock \emph{J. Low Temp. Phys.}, 131\penalty0 (3-4):\penalty0 401--412, May 2003.
\newblock URL \url{https://link.springer.com/article/10.1023/A:1022974413704}.

\bibitem[Pintschovius et~al.(2004)Pintschovius, Reznik, Reichardt, Endoh, Hiraka, Tranquada, Uchiyama, Masui, and Tajima]{Pintschovius2004}
L.~Pintschovius, D.~Reznik, W.~Reichardt, Y.~Endoh, H.~Hiraka, J.~M. Tranquada, H.~Uchiyama, T.~Masui, and S.~Tajima.
\newblock {Oxygen phonon branches in ${\mathrm{YBa}}_{2}{\mathrm{Cu}}_{3}{\mathrm{O}}_{7}$}.
\newblock \emph{Phys. Rev. B}, 69:\penalty0 214506, Jun 2004.
\newblock \doi{10.1103/PhysRevB.69.214506}.
\newblock URL \url{https://link.aps.org/doi/10.1103/PhysRevB.69.214506}.

\bibitem[Pintschovius(2005)]{Pintschovius2005}
Lothar Pintschovius.
\newblock Electron–phonon coupling effects explored by inelastic neutron scattering.
\newblock \emph{Phys. Status Solidi B}, 242\penalty0 (1):\penalty0 30--50, January 2005.
\newblock \doi{10.1002/pssb.200404951}.
\newblock URL \url{https://doi.org/10.1002/pssb.200404951}.

\bibitem[Reznik et~al.(2006)Reznik, Pintschovius, Ito, Iikubo, Sato, Goka, Fujita, Yamada, Gu, and Tranquada]{reznik2006electron}
Dmitri Reznik, L~Pintschovius, M~Ito, S~Iikubo, M~Sato, H~Goka, M~Fujita, K~Yamada, GD~Gu, and JM~Tranquada.
\newblock Electron--phonon coupling reflecting dynamic charge inhomogeneity in copper oxide superconductors.
\newblock \emph{Nature}, 440\penalty0 (7088):\penalty0 1170--1173, 2006.
\newblock URL \url{https://doi.org/10.1038/nature04704}.

\bibitem[Reznik(2010)]{Reznik2010}
Dmitry Reznik.
\newblock {Giant Electron-Phonon Anomaly in Doped La\textsubscript{2}CuO\textsubscript{4} and Other Cuprates}.
\newblock \emph{Adv. Condens. Matter Phys.}, page 523549, 2010.
\newblock \doi{10.1155/2010/523549}.
\newblock URL \url{https://doi.org/10.1155/2010/523549}.

\bibitem[Ghiringhelli et~al.(2012{\natexlab{a}})Ghiringhelli, Tacon, Minola, Blanco-Canosa, Mazzoli, Brookes, Luca, Frano, Hawthorn, He, Loew, Sala, Peets, Salluzzo, Schierle, Sutarto, Sawatzky, Weschke, Keimer, and Braicovich]{GhiringhelliCDW}
G.~Ghiringhelli, M.~Le Tacon, M.~Minola, S.~Blanco-Canosa, C.~Mazzoli, N.~Ben Brookes, G.~M.~De Luca, A.~Frano, D.~G. Hawthorn, F.~He, T.~Loew, M.~Moretti Sala, D.~C. Peets, M.~Salluzzo, E.~Schierle, R.~Sutarto, G.~A. Sawatzky, E.~Weschke, B.~Keimer, and L.~Braicovich.
\newblock {Long-Range Incommensurate Charge Fluctuations in (Y,Nd)Ba$_2$Cu$_3$O$_{6+x}$}.
\newblock \emph{Science}, 337\penalty0 (6096):\penalty0 821--825, 2012{\natexlab{a}}.
\newblock URL \url{https://www.science.org/doi/abs/10.1126/science.1223532}.

\bibitem[Chang et~al.(2012)Chang, Holmes, Mesot, Liang, Bonn, Hardy, Watenphul, et~al.]{chang2012direct}
J~Chang, AT~Holmes, J~Mesot, Ruixing Liang, DA~Bonn, WN~Hardy, A~Watenphul, et~al.
\newblock {Direct observation of competition between superconductivity and charge density wave order in YBa$_2$Cu$_3$O$_{6.67}$}.
\newblock \emph{Nat. Phys.}, 8\penalty0 (12):\penalty0 871--876, 2012.
\newblock URL \url{https://doi.org/10.1038/nphys2456}.

\bibitem[Comin et~al.(2014)Comin, Frano, Yoshida, Eisaki, Schierle, Weschke, Sutarto, He, et~al.]{comin2014charge}
R~Comin, A~Frano, Y~Yoshida, H~Eisaki, E~Schierle, E~Weschke, R~Sutarto, F~He, et~al.
\newblock {Charge order driven by Fermi-arc instability in Bi$_2$Sr$_{2-x}$La$_x$CuO$_{6+ \delta}$}.
\newblock \emph{Science}, 343\penalty0 (6169):\penalty0 390--392, 2014.
\newblock \doi{10.1126/science.1242996}.
\newblock URL \url{https://www.science.org/doi/abs/10.1126/science.1242996}.

\bibitem[Wu et~al.(2015)Wu, Mayaffre, Kr{\"a}mer, Horvati{\'c}, Berthier, Hardy, Liang, Bonn, and Julien]{wu2015incipient}
Tao Wu, Hadrien Mayaffre, Steffen Kr{\"a}mer, Mladen Horvati{\'c}, Claude Berthier, WN~Hardy, Ruixing Liang, DA~Bonn, and Marc-Henri Julien.
\newblock {Incipient charge order observed by NMR in the normal state of YBa$_2$Cu$_3$O$_y$}.
\newblock \emph{Nat. Commun.}, 6\penalty0 (1):\penalty0 6438, 2015.
\newblock URL \url{https://doi.org/10.1038/ncomms7438}.

\bibitem[Hayden and Tranquada(2024)]{hayden2024charge}
Stephen~M Hayden and John~M Tranquada.
\newblock Charge correlations in cuprate superconductors.
\newblock \emph{Annu. Rev. Condens. Matter Phys.}, 15, 2024.
\newblock URL \url{https://doi.org/10.1146/annurev-conmatphys-032922-094430}.

\bibitem[Arpaia et~al.(2019)Arpaia, Caprara, Fumagalli, De~Vecchi, Peng, Andersson, Betto, De~Luca, Brookes, Lombardi, Salluzzo, Braicovich, Di~Castro, Grilli, and Ghiringhelli]{ArpaiaUNO}
R.~Arpaia, S.~Caprara, R.~Fumagalli, G.~De~Vecchi, YY. Peng, E.~Andersson, D.~Betto, G.M. De~Luca, N.B. Brookes, F.~Lombardi, M~Salluzzo, L.~Braicovich, C.~Di~Castro, M.~Grilli, and G.~Ghiringhelli.
\newblock {Dynamical charge density fluctuations pervading the phase diagram of a Cu-based high Tc superconductor}.
\newblock \emph{Science}, 365\penalty0 (6456):\penalty0 906--910, 2019.
\newblock URL \url{https://www.science.org/doi/abs/10.1126/science.aav1315}.

\bibitem[Yu et~al.(2020)Yu, Tabis, Bialo, Yakhou, Brookes, Anderson, Tang, Yu, and Greven]{yu2020unusual}
Biqiong Yu, W~Tabis, I~Bialo, F~Yakhou, NB~Brookes, Z~Anderson, Y~Tang, G~Yu, and M~Greven.
\newblock {Unusual dynamic charge correlations in simple-tetragonal HgBa$_2$CuO$_{4+\delta}$}.
\newblock \emph{Phys. Rev. X}, 10\penalty0 (2):\penalty0 021059, 2020.
\newblock URL \url{https://doi.org/10.1103/PhysRevX.10.021059}.

\bibitem[Wang et~al.(2020{\natexlab{a}})Wang, Yu, Jing, Luo, Zeng, Li, Bialo, Bluschke, Tang, Freyermuth, et~al.]{wang2020doping}
Lichen Wang, Biqiong Yu, Ran Jing, Xiangpeng Luo, Junbang Zeng, Jiarui Li, Izabela Bialo, Martin Bluschke, Yang Tang, Jacob Freyermuth, et~al.
\newblock {Doping-dependent phonon anomaly and charge-order phenomena in the HgBa$_2$CuO$_{4+\delta}$ and HgBa$_2$CaCu$_2$O$_{6+\delta}$ superconductors}.
\newblock \emph{Phys. Rev. B}, 101\penalty0 (22):\penalty0 220509, 2020{\natexlab{a}}.
\newblock URL \url{https://doi.org/10.1103/PhysRevB.101.220509}.

\bibitem[Wang et~al.(2020{\natexlab{b}})Wang, Horio, Von~Arx, Shen, John~Mukkattukavil, Sassa, Ivashko, Matt, Pyon, Takayama, et~al.]{wang2020high}
Qisi Wang, M~Horio, K~Von~Arx, Y~Shen, D~John~Mukkattukavil, Y~Sassa, O~Ivashko, CE~Matt, S~Pyon, T~Takayama, et~al.
\newblock {High-temperature charge-stripe correlations in La$_{1.675}$Eu$_{0.2}$Sr$_{0.125}$CuO$_4$}.
\newblock \emph{Phys. Rev. Lett.}, 124\penalty0 (18):\penalty0 187002, 2020{\natexlab{b}}.
\newblock URL \url{https://doi.org/10.1103/PhysRevLett.124.187002}.

\bibitem[Arpaia and Ghiringhelli(2021)]{arpaia2021charge}
Riccardo Arpaia and Giacomo Ghiringhelli.
\newblock Charge order at high temperature in cuprate superconductors.
\newblock \emph{J. Phys. Soc. Jpn.}, 90\penalty0 (11):\penalty0 111005, 2021.
\newblock URL \url{https://journals.jps.jp/doi/abs/10.7566/JPSJ.90.111005}.

\bibitem[von Arx et~al.(2023)von Arx, Wang, Mustafi, Mazzone, Horio, Mukkattukavil, Pomjakushina, Pyon, Takayama, Takagi, et~al.]{von2023fate}
K~von Arx, Qisi Wang, S~Mustafi, DG~Mazzone, Masafumi Horio, D~John Mukkattukavil, E~Pomjakushina, S~Pyon, T~Takayama, H~Takagi, et~al.
\newblock Fate of charge order in overdoped {L}a-based cuprates.
\newblock \emph{npj Quantum Mater.}, 8\penalty0 (1):\penalty0 7, 2023.
\newblock URL \url{https://doi.org/10.1038/s41535-023-00539-w}.

\bibitem[Arpaia et~al.(2023)Arpaia, Martinelli, Sala, Caprara, Nag, Brookes, Camisa, Li, Gao, Zhou, et~al.]{NatCommRArpaia}
Riccardo Arpaia, Leonardo Martinelli, Marco~Moretti Sala, Sergio Caprara, Abhishek Nag, Nicholas~B Brookes, Pietro Camisa, Qizhi Li, Qiang Gao, Xingjiang Zhou, et~al.
\newblock Signature of quantum criticality in cuprates by charge density fluctuations.
\newblock \emph{Nat. Commun.}, 14\penalty0 (1):\penalty0 7198, 2023.
\newblock URL \url{https://doi.org/10.1038/s41467-023-42961-5}.

\bibitem[Braicovich et~al.(2020)Braicovich, Rossi, Fumagalli, Peng, Wang, Arpaia, Betto, De~Luca, Di~Castro, Kummer, et~al.]{BraicovichPhon}
Lucio Braicovich, Matteo Rossi, Roberto Fumagalli, Yingying Peng, Yan Wang, Riccardo Arpaia, Davide Betto, Gabriella~M De~Luca, Daniele Di~Castro, Kurt Kummer, et~al.
\newblock {Determining the electron-phonon coupling in superconducting cuprates by resonant inelastic x-ray scattering: Methods and results on Nd$_{1+x}$Ba$_{2-x}$Cu$_3$O$_{7- \delta}$}.
\newblock \emph{Phys. Rev. Research}, 2\penalty0 (2):\penalty0 023231, 2020.
\newblock URL \url{https://journals.aps.org/prresearch/abstract/10.1103/PhysRevResearch.2.023231}.

\bibitem[Bernhard et~al.(2001)Bernhard, Tallon, Blasius, Golnik, and Niedermayer]{bernhard2001anomalous}
C~Bernhard, JL~Tallon, Th~Blasius, A~Golnik, and Ch~Niedermayer.
\newblock {Anomalous peak in the superconducting condensate density of cuprate high-T$_c$ superconductors at a unique doping state}.
\newblock \emph{Phys. Rev. Lett.}, 86\penalty0 (8):\penalty0 1614, 2001.
\newblock URL \url{https://doi.org/10.1103/PhysRevLett.86.1614}.

\bibitem[Tallon et~al.(2003)Tallon, Loram, Cooper, Panagopoulos, and Bernhard]{tallon2003superfluid}
JL~Tallon, JW~Loram, JR~Cooper, Christos Panagopoulos, and Christian Bernhard.
\newblock {Superfluid density in cuprate high-T$_c$ superconductors: A new paradigm}.
\newblock \emph{Phys. Rev. B}, 68\penalty0 (18):\penalty0 180501, 2003.
\newblock URL \url{https://doi.org/10.1103/PhysRevB.68.180501}.

\bibitem[Tallon(2014)]{tallon2014thermodynamics}
Jeffery~L Tallon.
\newblock {Thermodynamics and Critical Current Density in High-$T_{\mathrm{c}}$ Superconductors}.
\newblock \emph{IEEE Trans. Appl. Supercond.}, 25\penalty0 (3):\penalty0 1--6, 2014.
\newblock URL \url{https://doi.org/10.1109/TASC.2014.2379660}.

\bibitem[Grissonnanche et~al.(2014)Grissonnanche, Cyr-Choini{\`e}re, Lalibert{\'e}, Ren{\'e}~de Cotret, Juneau-Fecteau, Dufour-Beaus{\'e}jour, Delage, LeBoeuf, Chang, Ramshaw, et~al.]{grissonnanche2014direct}
G~Grissonnanche, O~Cyr-Choini{\`e}re, Fs~Lalibert{\'e}, S~Ren{\'e}~de Cotret, A~Juneau-Fecteau, S~Dufour-Beaus{\'e}jour, M-E Delage, D~LeBoeuf, J~Chang, BJ~Ramshaw, et~al.
\newblock Direct measurement of the upper critical field in cuprate superconductors.
\newblock \emph{Nat. Commun.}, 5\penalty0 (1):\penalty0 3280, 2014.
\newblock URL \url{https://doi.org/10.1038/ncomms4280}.

\bibitem[Yu et~al.(2024)Yu, Ciccarino, Bianco, Errea, Narang, and Bernevig]{yu2024non}
Jiabin Yu, Christopher~J Ciccarino, Raffaello Bianco, Ion Errea, Prineha Narang, and B~Andrei Bernevig.
\newblock Non-trivial quantum geometry and the strength of electron--phonon coupling.
\newblock \emph{Nat. Phys.}, 20\penalty0 (8):\penalty0 1262--1268, 2024.

\bibitem[Yu et~al.(2025)Yu, Bernevig, Queiroz, Rossi, T{\"o}rm{\"a}, and Yang]{yu2025quantum}
Jiabin Yu, B~Andrei Bernevig, Raquel Queiroz, Enrico Rossi, P{\"a}ivi T{\"o}rm{\"a}, and Bohm-Jung Yang.
\newblock Quantum geometry in quantum materials.
\newblock \emph{npj Quantum Mater.}, 10\penalty0 (1):\penalty0 101, 2025.

\bibitem[Julku et~al.(2021)Julku, Bruun, and T{\"o}rm{\"a}]{julku2021quantum}
Aleksi Julku, Georg~M Bruun, and P{\"a}ivi T{\"o}rm{\"a}.
\newblock Quantum geometry and flat band bose-einstein condensation.
\newblock \emph{Phys. Rev. Lett.}, 127\penalty0 (17):\penalty0 170404, 2021.

\bibitem[Arpaia et~al.(2018)Arpaia, Andersson, Trabaldo, Bauch, and Lombardi]{ArpaiaGrowth}
Riccardo Arpaia, Eric Andersson, Edoardo Trabaldo, Thilo Bauch, and Floriana Lombardi.
\newblock {Probing the phase diagram of cuprates with YBa$_2$Cu$_3$O$_{7-\delta}$ thin films and nanowires}.
\newblock \emph{Phys. Rev. Mater.}, 2:\penalty0 024804, Feb 2018.
\newblock \doi{10.1103/PhysRevMaterials.2.024804}.
\newblock URL \url{https://link.aps.org/doi/10.1103/PhysRevMaterials.2.024804}.

\bibitem[Sala et~al.(2011)Sala, Bisogni, Aruta, Balestrino, Berger, Brookes, De~Luca, Di~Castro, Grioni, Guarise, et~al.]{Moretti}
M~Moretti Sala, Valentina Bisogni, C~Aruta, G~Balestrino, H~Berger, N~Ben Brookes, GM~De~Luca, D~Di~Castro, M~Grioni, M~Guarise, et~al.
\newblock {Energy and symmetry of dd excitations in undoped layered cuprates measured by Cu L$_3$ resonant inelastic x-ray scattering}.
\newblock \emph{New J. Phys.}, 13\penalty0 (4):\penalty0 043026, 2011.
\newblock URL \url{https://iopscience.iop.org/article/10.1088/1367-2630/13/4/043026/meta}.

\bibitem[Minola et~al.(2015)Minola, Dellea, Gretarsson, Peng, Lu, Porras, Loew, Yakhou, Brookes, Huang, Pelliciari, Schmitt, Ghiringhelli, Keimer, Braicovich, and Le~Tacon]{Minola}
M.~Minola, G.~Dellea, H.~Gretarsson, Y.~Y. Peng, Y.~Lu, J.~Porras, T.~Loew, F.~Yakhou, N.~B. Brookes, Y.~B. Huang, J.~Pelliciari, T.~Schmitt, G.~Ghiringhelli, B.~Keimer, L.~Braicovich, and M.~Le~Tacon.
\newblock Collective nature of spin excitations in superconducting cuprates probed by resonant inelastic x-ray scattering.
\newblock \emph{Phys. Rev. Lett.}, 114:\penalty0 217003, May 2015.
\newblock \doi{10.1103/PhysRevLett.114.217003}.
\newblock URL \url{https://link.aps.org/doi/10.1103/PhysRevLett.114.217003}.

\bibitem[Ghiringhelli et~al.(2012{\natexlab{b}})Ghiringhelli, Le~Tacon, Minola, Blanco-Canosa, Mazzoli, Brookes, De~Luca, Frano, Hawthorn, He, et~al.]{ghiringhelli2012long}
G~Ghiringhelli, M~Le~Tacon, Matteo Minola, S~Blanco-Canosa, Claudio Mazzoli, NB~Brookes, GM~De~Luca, A~Frano, DG~Hawthorn, F~He, et~al.
\newblock Long-range incommensurate charge fluctuations in (y, nd) ba2cu3o6+ x.
\newblock \emph{Science}, 337\penalty0 (6096):\penalty0 821--825, 2012{\natexlab{b}}.

\bibitem[Martinelli et~al.(2025)Martinelli, Bia\l{}o, Hong, Oppliger, Lin, Schaller, K\"uspert, Fischer, Kurosawa, Momono, Oda, Novikov, Khadiev, Weschke, Choi, Agrestini, Garcia-Fernandez, Zhou, Wang, and Chang]{Martinelli2024}
L.~Martinelli, I.~Bia\l{}o, X.~Hong, J.~Oppliger, C.~Lin, T.~Schaller, J.~K\"uspert, M.~H. Fischer, T.~Kurosawa, N.~Momono, M.~Oda, D.~V. Novikov, A.~Khadiev, E.~Weschke, J.~Choi, S.~Agrestini, M.~Garcia-Fernandez, Ke-Jin Zhou, Q.~Wang, and J.~Chang.
\newblock Decoupling of static and dynamic charge correlations revealed by uniaxial strain in a cuprate superconductor.
\newblock \emph{Phys. Rev. B}, 112:\penalty0 L041124, Jul 2025.
\newblock \doi{10.1103/n834-dkyh}.
\newblock URL \url{https://link.aps.org/doi/10.1103/n834-dkyh}.

\bibitem[Wahlberg et~al.(2021)Wahlberg, Arpaia, Seibold, Rossi, Fumagalli, Trabaldo, Brookes, Braicovich, Caprara, Gran, et~al.]{wahlberg2021restored}
Eric Wahlberg, Riccardo Arpaia, G{\"o}tz Seibold, Matteo Rossi, Roberto Fumagalli, Edoardo Trabaldo, Nicholas~B Brookes, Lucio Braicovich, Sergio Caprara, Ulf Gran, et~al.
\newblock {Restored strange metal phase through suppression of charge density waves in underdoped YBa$_2$Cu$_3$O$_{7-\delta}$}.
\newblock \emph{Science}, 373\penalty0 (6562):\penalty0 1506--1510, 2021.
\newblock URL \url{https://www.science.org/doi/full/10.1126/science.abc8372}.

\bibitem[Boyle et~al.(2021)Boyle, Walker, Ruiz, Schierle, Zhao, Boschini, Sutarto, Boyko, Moore, Tamura, et~al.]{boyle2021large}
TJ~Boyle, M~Walker, A~Ruiz, E~Schierle, Z~Zhao, F~Boschini, R~Sutarto, TD~Boyko, W~Moore, N~Tamura, et~al.
\newblock {Large response of charge stripes to uniaxial stress in La$_{1.475}$Nd$_{0.4}$Sr$_{0.125}$CuO$_4$}.
\newblock \emph{Phys. Rev. Research}, 3\penalty0 (2):\penalty0 L022004, 2021.
\newblock URL \url{https://doi.org/10.1103/PhysRevResearch.3.L022004}.

\bibitem[Castellani et~al.(1995)Castellani, Di~Castro, and Grilli]{Castellani}
C.~Castellani, C.~Di~Castro, and M.~Grilli.
\newblock {Singular Quasiparticle Scattering in the Proximity of Charge Instabilities}.
\newblock \emph{Phys. Rev. Lett.}, 75:\penalty0 4650--4653, Dec 1995.
\newblock \doi{10.1103/PhysRevLett.75.4650}.
\newblock URL \url{https://link.aps.org/doi/10.1103/PhysRevLett.75.4650}.

\bibitem[Becca et~al.(1996)Becca, Tarquini, Grilli, and Di~Castro]{becca1996charge}
Federico Becca, M~Tarquini, Marco Grilli, and C~Di~Castro.
\newblock {Charge-density waves and superconductivity as an alternative to phase separation in the infinite-U Hubbard-Holstein model}.
\newblock \emph{Phys. Rev. B}, 54\penalty0 (17):\penalty0 12443, 1996.
\newblock URL \url{https://doi.org/10.1103/PhysRevB.54.12443}.

\bibitem[Blackburn et~al.(2013)Blackburn, Chang, Said, Leu, Liang, Bonn, Hardy, Forgan, and Hayden]{blackburn2013inelastic}
Elizabeth Blackburn, Johan Chang, AH~Said, BM~Leu, Ruixing Liang, Doug~A Bonn, WN~Hardy, Edward~M Forgan, and Stephen~M Hayden.
\newblock {Inelastic x-ray study of phonon broadening and charge-density wave formation in ortho-II-ordered YBa$_2$Cu$_3$O$_{6.54}$}.
\newblock \emph{Phys. Rev. B}, 88\penalty0 (5):\penalty0 054506, 2013.
\newblock URL \url{https://doi.org/10.1103/PhysRevB.88.054506}.

\bibitem[Chen et~al.(2016)Chen, Thampy, Mazzoli, Barbour, Miao, Gu, Cao, Tranquada, Dean, and Wilkins]{chen2016remarkable}
XM~Chen, V~Thampy, C~Mazzoli, AM~Barbour, H~Miao, GD~Gu, Y~Cao, JM~Tranquada, MPM Dean, and SB~Wilkins.
\newblock {Remarkable stability of charge density wave order in La$_{1.875}$Ba$_{0.125}$CuO$_4$}.
\newblock \emph{Phys. Rev. Lett.}, 117\penalty0 (16):\penalty0 167001, 2016.
\newblock URL \url{https://doi.org/10.1103/PhysRevLett.117.167001}.

\bibitem[Thampy et~al.(2017)Thampy, Chen, Cao, Mazzoli, Barbour, Hu, Miao, Fabbris, Zhong, Gu, et~al.]{thampy2017static}
V~Thampy, XM~Chen, Y~Cao, C~Mazzoli, AM~Barbour, W~Hu, H~Miao, G~Fabbris, RD~Zhong, GD~Gu, et~al.
\newblock {Static charge-density-wave order in the superconducting state of La$_{2-x}$Ba$_x$CuO$_4$}.
\newblock \emph{Phys. Rev. B}, 95\penalty0 (24):\penalty0 241111, 2017.
\newblock URL \url{https://doi.org/10.1103/PhysRevB.95.241111}.

\bibitem[Perali et~al.(1996)Perali, Castellani, Di~Castro, and Grilli]{perali1996d}
A~Perali, C~Castellani, C~Di~Castro, and M~Grilli.
\newblock d-wave superconductivity near charge instabilities.
\newblock \emph{Phys. Rev. B}, 54\penalty0 (22):\penalty0 16216--16225, 1996.
\newblock URL \url{https://doi.org/10.1103/PhysRevB.54.16216}.

\bibitem[Gweon et~al.(2004)Gweon, Sasagawa, Zhou, Graf, Takagi, Lee, and Lanzara]{gweon2004unusual}
G-H Gweon, T~Sasagawa, SY~Zhou, J~Graf, H~Takagi, D-H Lee, and A~Lanzara.
\newblock An unusual isotope effect in a high-transition-temperature superconductor.
\newblock \emph{Nature}, 430\penalty0 (6996):\penalty0 187--190, 2004.
\newblock URL \url{https://doi.org/10.1038/nature02731}.

\bibitem[Jiang et~al.(2024)Jiang, Ummarino, Baggioli, Liarokapis, and Zaccone]{jiang2024correlation}
Cunyuan Jiang, Giovanni~Alberto Ummarino, Matteo Baggioli, Efthymios Liarokapis, and Alessio Zaccone.
\newblock {Correlation between optical phonon softening and superconducting $T_\mathrm{c}$ in YBa$_2$Cu$_3$O$_x$ within d-wave Eliashberg theory}.
\newblock \emph{J. Phys. Mater.}, 7\penalty0 (4):\penalty0 045002, 2024.
\newblock URL \url{https://iopscience.iop.org/article/10.1088/2515-7639/ad6c7f/meta}.

\bibitem[Holt et~al.(2001)Holt, Zschack, Hong, Chou, and Chiang]{holt2001x}
Martin Holt, Paul Zschack, Hawoong Hong, MY~Chou, and T-C Chiang.
\newblock {X-ray studies of phonon softening in TiSe$_2$}.
\newblock \emph{Phys. Rev. Lett.}, 86\penalty0 (17):\penalty0 3799, 2001.
\newblock URL \url{https://doi.org/10.1103/PhysRevLett.86.3799}.

\bibitem[Monney et~al.(2016)Monney, Puppin, Nicholson, Hoesch, Chapman, Springate, Berger, Magrez, Cacho, Ernstorfer, et~al.]{monney2016revealing}
Claude Monney, Michele Puppin, CW~Nicholson, M~Hoesch, RT~Chapman, E~Springate, H~Berger, A~Magrez, C~Cacho, Ralph Ernstorfer, et~al.
\newblock {Revealing the role of electrons and phonons in the ultrafast recovery of charge density wave correlations in 1T-TiSe$_2$}.
\newblock \emph{Phys. Rev. B}, 94\penalty0 (16):\penalty0 165165, 2016.
\newblock URL \url{https://doi.org/10.1103/PhysRevB.94.165165}.

\bibitem[Chen et~al.(2022)Chen, Chen, Schnelle, Felser, and Gaulin]{chen2022charge}
Qiang Chen, D~Chen, W~Schnelle, C~Felser, and BD~Gaulin.
\newblock {Charge density wave order and fluctuations above $T_CDW$ and below superconducting $T_c$ in the kagome metal CsV$_{3}$Sb$_{5}$}.
\newblock \emph{Phys. Rev. Lett.}, 129\penalty0 (5):\penalty0 056401, 2022.
\newblock URL \url{https://doi.org/10.1103/PhysRevLett.129.056401}.

\bibitem[Cao et~al.(2023)Cao, Xu, Fukui, Manjo, Dong, Shi, Liu, Cao, and Song]{cao2023competing}
Saizheng Cao, Chenchao Xu, Hiroshi Fukui, Taishun Manjo, Ying Dong, Ming Shi, Yang Liu, Chao Cao, and Yu~Song.
\newblock {Competing charge-density wave instabilities in the kagome metal ScV$_6$Sn$_6$}.
\newblock \emph{Nat. Commun.}, 14\penalty0 (1):\penalty0 7671, 2023.
\newblock URL \url{https://doi.org/10.1038/s41467-023-43454-1}.

\bibitem[Liu et~al.(2024)Liu, Wang, Yao, Jia, Zhang, and Cho]{liu2024driving}
Shuyuan Liu, Chongze Wang, Shichang Yao, Yu~Jia, Zhenyu Zhang, and Jun-Hyung Cho.
\newblock {Driving mechanism and dynamic fluctuations of charge density waves in the kagome metal CsV$_{6}$Sb$_{6}$}.
\newblock \emph{Phys. Rev. B}, 109\penalty0 (12):\penalty0 L121103, 2024.
\newblock URL \url{https://doi.org/10.1103/PhysRevB.109.L121103}.

\bibitem[Guo et~al.(2025)Guo, Kogar, Henke, Flicker, de~Juan, Sun, Khayr, Peng, Lee, Krogstad, et~al.]{guo2025plane}
Xuefei Guo, Anshul Kogar, Jans Henke, Felix Flicker, Fernando de~Juan, Stella X-L Sun, Issam Khayr, Yingying Peng, Sangjun Lee, Matthew~J Krogstad, et~al.
\newblock {In-plane anisotropy of charge density wave fluctuations in 1$ T $-TiSe$_2$}.
\newblock \emph{Phys. Rev. Lett.}, 135:\penalty0 136102, 2025.
\newblock URL \url{https://doi.org/10.1103/j8vm-wb65}.

\bibitem[Fragkos et~al.(2025)Fragkos, Orio, Erhardt, Jabed, Sasi, Courtade, Masilamani, {\"U}nzelmann, Diekmann, Hildebrand, et~al.]{fragkos2025electron}
Sotirios Fragkos, Hibiki Orio, Nina~Girotto Erhardt, Akib Jabed, Sarath Sasi, Quentin Courtade, Muthu Masilamani, Maximilian {\"U}nzelmann, Florian Diekmann, Baptiste Hildebrand, et~al.
\newblock {Electron-phonon-dominated charge-density-wave fluctuations in TiSe$_2$ accessed by ultrafast nonequilibrium dynamics}.
\newblock \emph{arXiv preprint arXiv:2507.12430}, 2025.
\newblock URL \url{https://arxiv.org/abs/2507.12430}.

\bibitem[Liu et~al.(2025)Liu, Duan, Liu, Liu, Wang, Gu, Huang, Yang, Liu, Qian, et~al.]{liu2025fluctuated}
Haoran Liu, Shaofeng Duan, Xiangqi Liu, Zhihua Liu, Shichong Wang, Lingxiao Gu, Jiongyu Huang, Wenxuan Yang, Jianzhe Liu, Dong Qian, et~al.
\newblock {Fluctuated lattice-driven charge density wave far above the condensation temperature in kagome superconductor KV$_3$Sb$_5$}.
\newblock \emph{Sci. Bull.}, 2025.
\newblock URL \url{https://doi.org/10.1016/j.scib.2025.02.018}.

\bibitem[Liang et~al.(2006)Liang, Bonn, and Hardy]{liang2006evaluation}
Ruixing Liang, DA~Bonn, and WN~Hardy.
\newblock Evaluation of {C}u{O}$_2$ plane hole doping in {YB}a$_2${C}u$_3${O}$_{6+x}$ single crystals.
\newblock \emph{Phys. Rev. B}, 73\penalty0 (18):\penalty0 180505, 2006.
\newblock URL \url{https://doi.org/10.1103/PhysRevB.73.180505}.

\bibitem[Ament et~al.(2011)Ament, Van~Veenendaal, Devereaux, Hill, and Van Den~Brink]{ament2011resonant}
Luuk~JP Ament, Michel Van~Veenendaal, Thomas~P Devereaux, John~P Hill, and Jeroen Van Den~Brink.
\newblock Resonant inelastic x-ray scattering studies of elementary excitations.
\newblock \emph{Rev. Mod. Phys.}, 83\penalty0 (2):\penalty0 705--767, 2011.
\newblock URL \url{https://doi.org/10.1103/RevModPhys.83.705}.

\bibitem[Devereaux et~al.(2016)Devereaux, Shvaika, Wu, Wohlfeld, Jia, Wang, Moritz, Chaix, Lee, Shen, Ghiringhelli, and Braicovich]{DevereauxPhon}
T.~P. Devereaux, A.~M. Shvaika, K.~Wu, K.~Wohlfeld, C.~J. Jia, Y.~Wang, B.~Moritz, L.~Chaix, W.-S. Lee, Z.-X. Shen, G.~Ghiringhelli, and L.~Braicovich.
\newblock {Directly Characterizing the Relative Strength and Momentum Dependence of Electron-Phonon Coupling Using Resonant Inelastic X-Ray Scattering}.
\newblock \emph{Phys. Rev. X}, 6:\penalty0 041019, 2016.
\newblock URL \url{https://link.aps.org/doi/10.1103/PhysRevX.6.041019}.

\end{thebibliography}
\end{document}